\documentclass[13pt,letterpaper]{article}
\usepackage[margin=1in]{geometry}
\usepackage{amsmath,amssymb}

\usepackage[utf8]{inputenc}

\usepackage{nameref,hyperref}
\usepackage{amssymb}
\usepackage{amsmath,amsfonts}
\usepackage{algorithm}
\usepackage[noend]{algpseudocode} 
\usepackage{array}
\usepackage[caption=false,font=normalsize,labelfont=sf,textfont=sf]{subfig}
\usepackage{textcomp}
\usepackage{stfloats}
\usepackage{url}
\usepackage{verbatim}
\usepackage{booktabs}
\usepackage{graphicx}
\usepackage{hyperref}
\usepackage[dvipsnames]{xcolor}
\usepackage{cite}
\usepackage{dsfont}
\usepackage{tabularx}
\usepackage{amsmath}
\usepackage{xspace}
\usepackage{multirow} 
\usepackage{xcolor}
\usepackage{rotating}
\usepackage{subcaption}
\usepackage{listings}
\usepackage{adjustbox}

\usepackage{microtype}
\DisableLigatures[f]{encoding = *, family = * }

\usepackage{changepage}

\usepackage[aboveskip=1pt,labelfont=bf,labelsep=period,singlelinecheck=off]{caption}

\makeatletter
\renewcommand{\@biblabel}[1]{\quad#1.}
\makeatother

\usepackage{lastpage,fancyhdr,graphicx}
\usepackage{epstopdf}
\fancyheadoffset[L]{2.25in}
\fancyfootoffset[L]{2.25in}

\usepackage{color}

\definecolor{Gray}{gray}{.25}

\usepackage{graphicx}

\usepackage{sidecap}

\usepackage{wrapfig}
\usepackage[pscoord]{eso-pic}
\usepackage[fulladjust]{marginnote}
\reversemarginpar

\begin{document}
\vspace*{0.35in}

\begin{flushleft}
{\Large
\textbf\newline{{\em Trust@Health}: A Trust-Based Multilayered Network for Scalable Healthcare Service Management}
}
\newline
\\
Avijit Gayen\textsuperscript{1,3},
Somyajit Chakraborty\textsuperscript{2,*},
Joydeep Chakraborty\textsuperscript{3},
Angshuman Jana\textsuperscript{1}
\\
\bigskip
\textbf{1} Department of Computer Science, Indian Institute of Information Technology, Guwahati, India
\\
\textbf{2} Department of Chemical Engineering, Shanghai Jiao Tong University, China
\\
\textbf{3} Department of Computer Science and Engineering, Techno India University
\\
\bigskip
* chksomyajit@sjtu.edu.cn

\end{flushleft}

\section*{Abstract}
We study the intricate relationships within healthcare systems, focusing on interactions among doctors, departments, and hospitals. Leveraging an evolutionary graph framework, the proposed model emphasizes both intra-layer and inter-layer trust relationships to better understand and optimize healthcare services. The trust-based network facilitates the identification of key healthcare entities by integrating their social and professional interactions, culminating in a trust-based algorithm that quantifies the importance of these entities. Validation with a real-world dataset reveals a strong correlation (0.91) between the proposed trust measures and the ratings of hospitals and departments, though doctor ratings demonstrate skewed distributions due to potential biases. By modeling these relationships and trust dynamics, the framework supports scalable healthcare infrastructure, enabling effective patient referrals, personalized recommendations, and enhanced decision-making pathways.

\section{Introduction}\label{s:intro}Access to modern healthcare services in rural India and other countries worldwide is severely limited~\cite{rao2013rural}. As a consequence, it triggers the alarming practice of `self-medication', the purchase of medicine without a prescription, etc. The lack of availability of information regarding specialist doctors, hospitals, etc., also hinders access to medical services in the critical situations of patients~\cite {ahmad2014evaluation}. In a healthcare system, physicians rely on their relationships with physician colleagues for patient referrals, clinical advice, and information about the latest clinical advances \cite{guo2016doctor}.   However, the healthcare system is a part of the social system, and it is represented using a social network in various existing studies \cite{gao2014developing,barnett2011mapping}. 
In a recent study \cite{guo2016doctor}, Guo et al. proposed a `patient-centric' feed-forward model that incorporates the interconnectivity of individual medical entities like doctors, hospitals, and departments. This model can capture the interactive role of various medical entities in the healthcare system to facilitate medical services. Over the past, in social networks, social interactions have been portrayed by friendship, closeness, trust, partnerships, and many more relationships \cite{aghdam2020uncertainty}. However, recent studies have shown more interest in developing the trust-based model of social systems \cite{frankel2019physicians,abou2020ditrust,gao2014developing}.
\par Trust plays a crucial role in social interactions among the entities in any social system~\cite{mikucka2017does,bargain2020trust,nunkoo2017governance}. The existence of trust in various social systems has been one of the main interests of research of social scientists for a long time~\cite {missimer2017strategic}.  In literature~\cite{scheer2012trust}, trust has been described as an aggregate of several incentives of social interactions. It may incorporate several distinct concepts like convergence of interests, compatibility of incentives, competence, and knowledge. There exists a plethora of works where trust is used in the development of various socio-economic applications~\cite{o2005trust,artz2007survey,gayen2017towards}.


\par In literature,  many research works  \cite{frankel2019physicians,abou2020ditrust,gao2014developing,kethers2005modelling} have been proposed where trust plays a vital role in medical or healthcare networks. However, all the existing works\cite{harris2021covid,lowry2014understanding,larosa2018impacts} of healthcare systems that observe the existence of trust in the healthcare system are unable to provide a complete framework to capture the social interactions among the several existing medical entities. In a recent work~\cite{mondal2020building}, Mondal et al. proposed a multi-layered trust-based model to capture the patient-doctor interactions on a temporal basis. They modeled the interaction episode as a different layer to develop a trust-based doctor recommendation system.  However, they ignore the role of other essential healthcare entities in their model.
On the other hand, it is observed that the proposed model by Guo et al.~\cite{guo2016doctor} neglects the social aspect of the interactions among the individual medical entities. Moreover, in some recent studies~\cite{gopichandran2013factors, guo2016doctor}, it is observed that patients tend to choose doctors based on the number of workplaces or hospitals they have worked in and also the number of departments they are associated with. The nature of activities in the healthcare system necessitates a detailed study of interactions among the various medical entities to obtain the best possible solutions. Therefore, there is a strong requirement for a suitable network model based on the social interactions among the various medical entities. Furthermore, this study addresses the gap in managing and optimizing multilayered healthcare networks, offering insights into trust-driven service management applicable to scalable network architectures. 
In the literature,  many research works have been proposed. Existing works on healthcare systems, such as \cite{harris2021covid,larosa2018impacts}, did not consider the multilayered network designed to emphasize discrete relationships among the fundamental entities that constitute the backbone of any healthcare system – Doctors, Departments, and Hospitals.
\par Trust evaluation in healthcare networks is a crucial factor in ensuring effective collaboration among medical professionals and institutions. Our previous study, Ml-HCN~\cite{gayen2024ml}, introduced a multilayered framework for modeling healthcare entities, incorporating departmental, institutional, and professional trust factors. However, this model did not sufficiently capture real-time fluctuations in trust relationships, leading to potential inaccuracies in trust-based recommendations. To address this limitation, we propose an improved trust estimation algorithm, refining the methodology introduced in ~\cite{gayen2017towards} by integrating adaptive weighting and dynamic node interactions. \par Based on the proposed multi-layered network, we further develop a trust network to capture the trust relationship among the medical entities, which influences any social interactions among the entities in a social system. In our proposed model, we specifically consider the relationship among three important medical entities -- doctors, departments, and hospitals in healthcare systems as a multi-layered network. In our work, we introduce the concept of social score, which is an aggregated social importance of any medial entity in the proposed multi-layered network. We specifically use the \textit{`page-rank'} kind of algorithm to estimate the social score. This aggregated score not only considers the social interactions among the homogeneous entity, e.g., doctor-doctor interactions, department-department interactions, etc, but it is also estimated based on the interactions among heterogeneous entities, e.g., doctor-department interactions, department-hospital interactions, etc. Though the social score has been introduced in many social applications, including financial systems to measure the creditworthiness of any financial entity~\cite{oskarsdottir2019value}, academic social systems to estimate the educational impact of a researcher~\cite{gayen2017towards}, etc., we hardly find any contributory work where the social score has been introduced in the medical healthcare system. The social score of the medical entity in the medical \& healthcare system would not only help in distinguishing important doctors, hospitals, etc. but also provide the platform for the development of recommendation systems, chatbot-based patient interactive support systems, etc., in the healthcare domain. Finally, we perform various experiments to measure the performance of our proposed model. We initially do simulations to validate the proposed model. Further, the empirical analysis of the dataset reveals the nature of the social score and its usefulness. The patient is an integral part of any healthcare system. Therefore, in some earlier works, patient-doctor interactions have been separately modeled. Although we propose a multi-layered network for the healthcare system, the modeling of patient-patient social interaction might increase the complexity of the proposed model and be insignificant from our proposed perspectives. Thus, in the model, we considered the patient to be an external entity in healthcare. 
\par Specifically, the multi-layered healthcare model leverages an evolutionary graph system, emphasizing discrete relationships among the fundamental entities that constitute the backbone of any healthcare system – Doctors, Departments, and Hospitals.
Further, we introduce a mathematical model to formalize the social interactions among the medical entities. Observe that the proposed multilayered network model in healthcare can be utilized to streamline and improve the referral process, ensuring that patients are directed to the most suitable specialists or departments within suitable hospitals. The multilayered approach allows for a more nuanced evaluation, considering not only the individual expertise of doctors but also the collaborative dynamics between departments and the overall reputation of hospitals. 
Therefore, the model can enhance decision-making processes within healthcare systems, leading to more efficient and reliable patient care pathways.
While patient-patient interactions are valuable in contexts like social support networks, the scope of this work is deliberately restricted to institutional entities (doctors, departments, hospitals) to maintain computational tractability, data reliability, and alignment with operational healthcare goals. This design choice is consistent with prior literature and enables focused optimization of referral pathways and resource allocation. Future iterations may integrate patient networks as external modules, but their inclusion in the core model would dilute its primary objectives without commensurate benefits. It is important to note that the scope of this work lies in healthcare service management and network-based decision support rather than in biomedical engineering or clinical treatment validation. The proposed framework models structural, relational, and trust-based interactions among healthcare entities to enhance service optimization, referral pathways, and resource allocation. As such, it does not require clinical trials, medical diagnostics, or direct patient-level interventions, but can be adapted to operational contexts in collaboration with healthcare providers.
The contributions are summarized as follows:
\begin{itemize}
    \item We define a multi-layered medical network to model the social interactions among the various medical entities, and we formalize the social interactions using the relational model.
    \item We establish a trust-network model to capture the role of social interactions among the various medical entities in the healthcare system under trust.
    \item We introduce the social score of each medical entity based on the social trust relationship amongst the different medical entities in the healthcare system.  
    \item A novel algorithmic modification that enhances trust propagation dynamics, mitigating the impact of biased or skewed data distributions. It ensures the scalability and improved accuracy in real-world healthcare applications.
    %
    \item Finally, we simulate our proposed multi-layered network model using the collected data set to validate it, and we provide a case study.
    \item The comprehensive experimental validation of our proposed model demonstrates superior performance in comparison with both Ml-HCN~\cite{gayen2024ml} and the earlier work~\cite{gayen2017towards} on social trust.
\end{itemize}

\par We organize the paper as follows: Section \ref{s:related} discusses the detailed literature associated with our work to identify the scope of the work. In Section~\ref{s:proposed}, we describe the proposed multi-layered network model for the healthcare system. Further, we formalize the relationship among the entities of the proposed system. We also introduced the concept of trust network further in that section. In section~\ref{s:result}, we describe the dataset used for empirical analysis. We further describe the detailed outcome of the study of the dataset and interpret it from the perspective of this work. Finally, we conclude in section~\ref{s:conc} and put the remarks regarding the future direction of the work. 

\section{Related Work}\label{s:related}In this section, we outline the works related to healthcare and medical service care systems. Initially, we highlight some works related to healthcare systems and various behavioral activities of several medical entities in the concerned system. Further, we outline various works related to network models that incorporate the social interactions among those entities in the healthcare system. We also highlight some important related works that observe social interactions from the perspective of trust. Finally, we outline some recent works related to the application of social trust, where social trust-based scores have been used. 
\par{\bf Healthcare System \& behavioral activities: } The study of the healthcare system to understand the underlying structure and behavior of several entities in it was a strong interest of researchers from the early age of science\cite{wendt2009healthcare,wendt2014changing,wu2010overview}. The referral process is one of the common phenomena of the medical \& healthcare system.
A referral-based network study~\cite{iyengar2011opinion,christakis2011commentary} shows that doctors refer the patients within their personal networks. 
In another earlier work~\cite{gonzalez1991physician}, authors observe that various factors influence the referral system in the medical and healthcare systems. One of the major factors is workplace influence, i.e., hospitals. It plays an important role in the referral system.  In some other studies~\cite{keating1998physicians, gabbay2004evidence}, authors observed that doctors rely on their personal network among other doctors to seek clinical advice and advanced clinical information. 
A study from another aspect of the healthcare system, the authors revealed that the popularity of the doctor influences the doctor selection process by the patient, measured in terms of the number of associations with the hospitals~\cite{gopichandran2013factors}. These observations figure out the need for a detailed study of the interactions not only among the medical practitioners but also other medical entities e.g., hospitals and departments.   
\par{\bf Network model of Healthcare systems:} In pursuance of the strong requirement of a detailed study of interactions among the several medical entities, people further observed the system as a network to model the interactions in it more systematically. In a work~\cite{barnett2011mapping}, Barnett et al. proposed a mapping for the network of physicians with the help of administrative data using a small-world network analysis. The study provided a detailed, complex network of physician networks. In another work~\cite{hewett2013trust}, the authors confirmed that sharing patients is indeed a key factor influencing the medical doctor networks. 
In this work, the authors observe that departments are also key players in a medical network. 
Some earlier works~\cite{chambers2012social,keating2007factors} addressed the issue of lack of data in the study of medical networks and have shown, using Social Network Analysis(SNA) and ego mining studies, that social media can be used to negate the problem of lack of data about physicians or hospitals. 
\par The detailed overall network of medical entities like hospitals, doctors, and departments was presented by Guo et al. in~\cite{guo2016doctor}, which provided the groundwork for the establishment of any complex network of these said entities. Their study was based primarily on an opinion leader-based study~\cite{sharara2011active}, which made their study include only the key doctors, instead of all the doctors in any medical healthcare system. They did not explore the social aspect in their network modeling, specifically in their work, as they primarily concentrated on establishing a recommender system.
\par{\bf Trust in Healthcare systems:} The earlier studies~\cite{keating1998physicians, gabbay2004evidence} reveal that overall approval of informal consultation among physicians is strongly associated with beliefs about how it affects quality of care. This could be strong evidence of the need to explore healthcare networks from a trust-based social aspect. Though trust networks in healthcare services are relatively new, there are considerable foundation works~\cite{boyer2010social,cunningham2012health,montgomery2019burnout} in this field. Recent literature underscores the increasing use of social network analysis (SNA) techniques for process improvement within healthcare systems—highlighting the relevance of modeling trust and relational dynamics in operational settings~\cite{francis2024exploring}. Traditionally, {\em trust} plays a key role in assessing health care services, more specifically as a performance indicator \cite{davies1998trust}. The first apt comparative study on the role of trust affecting the physician-physician relationship as well as the patient-physician relationship was studied in the early 2000s by Pearson and Raeke~\cite{pearson2000patients}. Their work incorporated a synopsis of theories about patient trust and the evolution of methods to measure it. Another earlier work~\cite{calnan2004public}, includes the role of public trust in assessing the expertise of medical professionals. In a recent work~ \cite{abou2020ditrust}, authors proposed the concept of DITrust Chain that incorporates blockchain for sustainable healthcare IoT systems. Although it ensures secure and reliable data exchange, it is limited to IoT applications. It does not consider the broader inter-entity relationships, e.g., the relationship among doctors, departments, and hospitals. In another recent work~\cite{Zhuetal2024}, a zero-trust blockchain-enabled framework was introduced aimed at securing next-generation healthcare communication networks through context-aware and dynamic authentication mechanisms. Although their approach effectively enhances security, it primarily focuses on communication aspects without fully addressing scalability and the propagation of trust across multiple layers of healthcare entities.
\par
\begin{figure*}
    \includegraphics[width=\textwidth, height=5 cm]{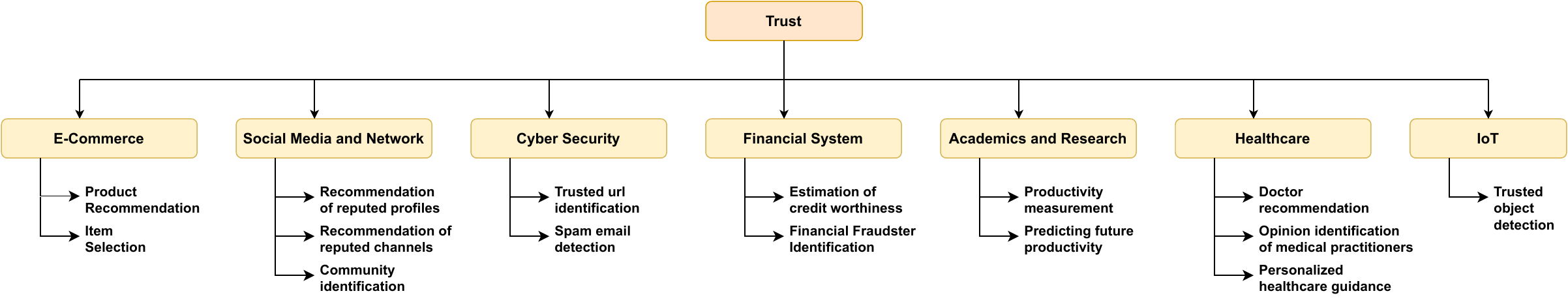}
    \caption{Figure represents a schematic tree structure to show the several domains of application of trust. The main domains are represented in a round rectangular box and the various important topics are shown below.}
    \label{fig:Applicat_trust}
\end{figure*}

 Though trust-based interactions are explored in many social systems where a composite social trust value of the entities has been estimated, we find very few works in the healthcare domain where the social trust score of the healthcare entities is estimated. Our work addresses this gap by introducing a multilayered trust-driven model that integrates interactions across doctors, departments, and hospitals, enabling scalability and adaptability for healthcare service optimization. Earlier in some work~\cite{oskarsdottir2019value}, we find that creditworthiness has been measured based on the social trust-based score. In another work~\cite{gayen2017towards}, Gayen et al. observed that a trust-based score could be a good indicator of the academic productivity of the researchers and it could be used as a predictor of their future impact in their academic domain. In figure~\ref{fig:Applicat_trust}, we represent the various applications of trust in several domains.

\par {\bf Multi-layered Network Model:} We recognize the emerging challenges in multilayer network models, particularly in terms of scalability, operator support, and software implementation. In recent studies, it focused on general-purpose systems too~\cite{panayiotou2024current}. In a recent work~\cite{mondal2020building}, Mondal et al. estimated the social trust score of the doctors based on the temporal model of patient-doctor interactions for doctor recommendation, though they failed to capture the patient referral among the doctors, collaborations among the hospitals, etc. which are very frequent event in the healthcare domain. In another earlier work~\cite{guo2016doctor}, A feed-forward model of the medical healthcare system was proposed by Guo et al. in 2016, which consisted of a `patient-centered' model with a network study for each individual medical entity, like hospitals, doctors, and their departments \cite{guo2016doctor}. These models failed to capture the social interrelation among these medical entities, which a trust-based multilayered network model could efficiently explore.
\par Prior works in healthcare trust modeling have primarily focused on single-layered or static evaluations of trust. Our earlier study, Ml-HCN~\cite{gayen2024ml}, introduced a multilayered network structure, providing a more comprehensive representation of trust relationships. However, it lacked a real-time adaptation mechanism, limiting its applicability in dynamic healthcare environments. Thus, there is a strong requirement for a trust-based healthcare framework that can capture the interactions not only between the doctor and patient but also the interactions among the doctor, department, and hospital. However, our proposed model not only addresses the intra-dependability among the entities, e.g., doctor-doctor, hospital-hospital, etc., but also the inter-dependability among medical entities like doctors, hospitals, and departments.  Additionally, in our previous research on Social Trust~\cite{gayen2017towards}, we developed a trust-based evaluation metric for scientific productivity, which, while effective, exhibited challenges in handling evolving trust dynamics and data sparsity. Our present work builds upon these foundations, introducing a modified trust estimation algorithm.

\section{Proposed Method} \label{s:proposed}
In our proposed network model, we specifically consider three healthcare entities, i.e., doctors, hospitals, and the various departments of the hospitals. We specifically propose a multi-layered network model to represent systematically the existing healthcare system. We model the relations among the medical entities as a trusted network to explore the interpersonal activities amongst them and compute the social score of the healthcare entities. Finally, we simulate the network and compute the experimental results. To capture interaction among the entities, we define the multi-layered network model below.
\subsection{Multi-layered network model}\label{subsec:medical}In this proposed network, we consider the entities in the healthcare system as the nodes in the network, and the edge represents the interactions in terms of similarity features, `belong to', etc., among the nodes. Our proposed multi-layered healthcare system broadly consists of two different parts, i.e., intra-layer networks and inter-layer networks. In table~\ref{tab:notation}, we depict several commonly used symbols in our work to describe the proposed model.
\par 
We define the multi-layer network $\mathbf{M}=(\emph{\textbf{G}}, \mathbf{I})$, where $\emph{\textbf{G}}$ represents the network and $\mathbf{I} \in \mathbb{R}$ is the number of layers. The network $\emph{\textbf{G}}$ is decomposed into several layer networks $\emph{\textbf{G}} = \{G_1, G_2, \dots, G_i\}$ where $i \in \mathbf{I}$. As an example, let us consider $\mathbf{M} = (\emph{\textbf{G}}, 3)$ and $\emph{\textbf{G}} = \{G_1, G_2, G_3\}$, where $G_1$ denotes the hospital network, $G_2$ denotes the departmental network, and $G_3$ denotes the doctors network.

\begin{table}[ht]
    \caption{List of Notations}
    \centering
    \renewcommand{\arraystretch}{1.2} 
    \begin{tabular}{|c|p{2.5in}|}
    \hline
    \textbf{Symbol} & \textbf{Description} \\ \hline
    $\mathbf{M}$ & The multi-layer network \\ \hline
    $X_\alpha$ & Set of nodes of the layer $\alpha$ network \\ \hline
    $E_\alpha$ & Set of edges of the layer $\alpha$ network \\ \hline
    $X_{i}^\alpha$ & Set of parameters of node $X_{i}^\alpha$ \\ \hline
    $W_\alpha$ & Weight of edges in the layer $\alpha$ network \\ \hline
    $P_\alpha$ & Total set (union) of parameters of nodes in the layer $\alpha$ network \\ \hline
    $W_{ij}^\alpha$ & Weight of the edge between $i$ and $j$ in the layer $\alpha$ network \\ \hline
    $A^{[\alpha]}$ & Adjacency matrix of nodes in the layer $\alpha$ network \\ \hline
    $E_{ij}^{\alpha\beta}$ & Inter-layer edge between node $i$ in the layer $\alpha$ network and node $j$ in the layer $\beta$ network \\ \hline
    $W_{ij}^{\alpha\beta}$ & Inter-layer edge weight between node $i$ in the layer $\alpha$ network and node $j$ in the layer $\beta$ network \\ \hline
    $A^{[\alpha,\beta]}$ & Inter-layer adjacency matrix of nodes in the layers $\alpha$ and $\beta$ networks \\ \hline
    \end{tabular}
    \label{tab:notation}
\end{table}


\subsubsection{Intra-layer network}In this section, we define three intra-layer networks, i.e., hospital ($G_1$), department ($G_2$), and doctor ($G_3$).
To create an intra-layer network, entities in that layer are considered as nodes of a network, and connections between two nodes are defined by the count of similar attributes, e.g., department, doctor, and hospital for $G_1$, $G_2$, $G_3$, respectively, that exist between any two nodes.

Let $n_1, n_2 \in G_i$ be two random nodes. Let $P_{\alpha} = \{\alpha_1, \alpha_2, \dots, \alpha_n\}$ be the set of attributes in $n_1$ and $P_{\beta} = \{\beta_1, \beta_2, \dots, \beta_n\}$ be the set of attributes in $n_2$. Then the similarity $\mathds{S}$ between $n_1$ and $n_2$ is defined as:

\[
\mathds{S}(n_1, n_2) = \frac{|P_{\alpha} \cap P_{\beta}|}{|P_{\alpha} \cup P_{\beta}|}
\]
\begin{equation}
\mathds{S} =  | P_{ \alpha } \cap P_{\beta} |~\mbox{ where}~  ~P_{ \alpha }~\mbox{and}~P_{\beta} \notin G_i
\label{eq:genral}
\end{equation}
\begin{figure}[!htb]
    \centering
    \includegraphics[width=0.5\textwidth, height=10cm ]{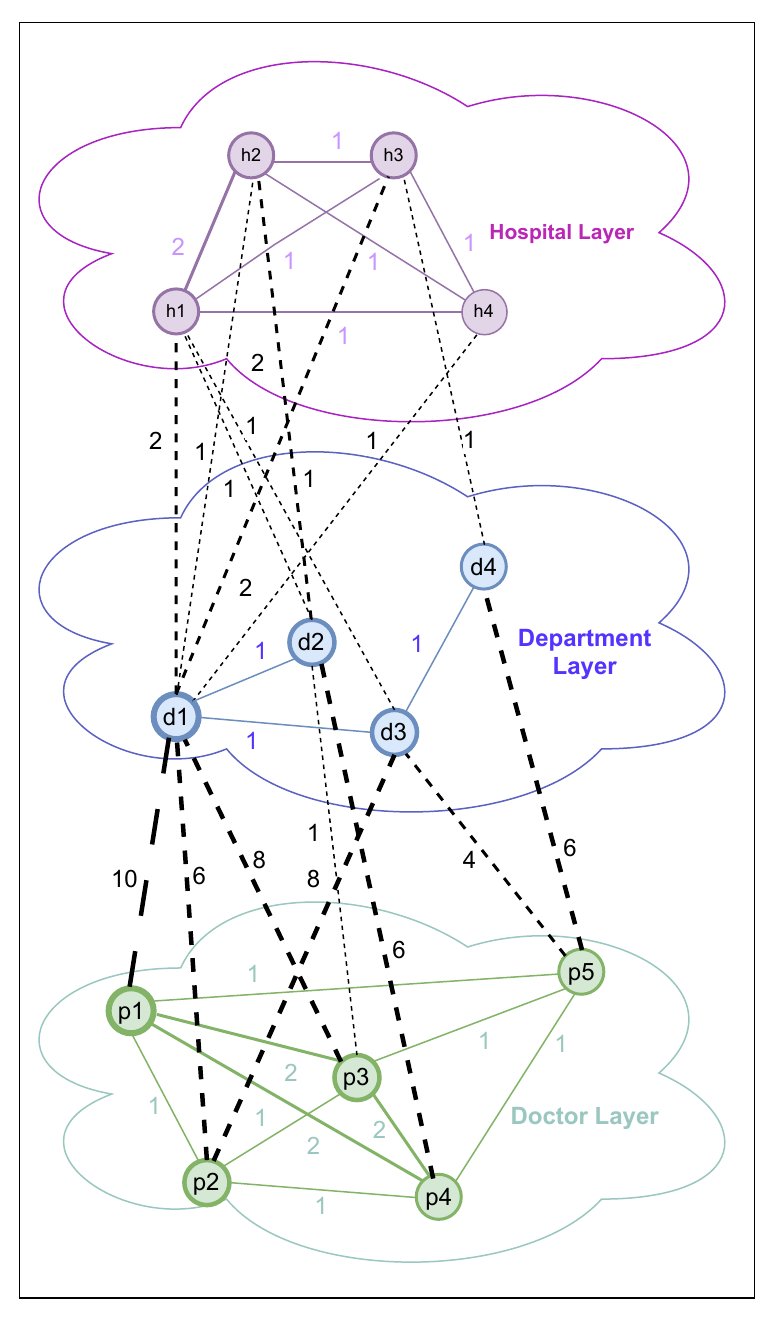}
    \caption{A schematic diagram of a Multi-layered Network. It shows three layers: hospital, department, and doctor layers in three different colors. Edges are also colored the same as the nodes of the corresponding layer. Inter-layer edges are black-colored dashed lines. Bold edges denote high edge weight.}
    \label{fig:toy_network}
\end{figure}

\subsubsection{Inter-layer Network}
In this section, we establish the inter-layer edges amongst the layers $G_1$ and $G_2$; on the other hand, $G_2$  and $G_3$ capture the `belongs to' relationship between the entities of different layers. Though there exist the inter-layer edges amongst the layers $G_1$ and $G_3$ logically, as this relationship is recursively computed, we ignore this relationship in our network. The weight of the edges in the inter-layered network is based on the importance of the node in the `belongs to' relationship which can be captured by several features of the node, e.g., the importance of a department in a hospital can be computed by the count of the doctors in that department, the number of special equipment items available in the concerned department, etc. Similarly, the importance of a doctor in a department in a specific hospital can be measured by tenure of association and his/her qualifications, etc.
\par In our model, departmental importance metrics (e.g., doctor count) are inherently normalized through the trust propagation algorithm (discussed in section~\ref{s:trust_net}), where edge weights are scaled by their row sums. This ensures comparability across heterogeneous metrics. For example, a department with 10 doctors in a hospital of 100 total doctors contributes $W_{ij} = 0.1$ (weight between node $i$ in the hospital
network and node $j$ in the department network) to the inter-layer adjacency matrix $A^{[hd]}$. Similarly, equipment availability is scaled by a predefined priority score (e.g., MRI machines are weighted higher than standard tools).

\subsection{Formalization of Relations}\label{subsec:relational}In this section, we propose the relational theorem to capture the interactions among several medical entities.
\subsubsection{Intra-layer relationship} In this section, we propose the relational theorem of the intra-layer network. Initially, we propose a general theorem for each layer of the network. We further use a weighted adjacency matrix to capture the overall view of each network.
\paragraph*{Definition 1}
In a layer \(G_\alpha=(X_\alpha, E_\alpha)\), where \(X_\alpha\) are the nodes and \(E_\alpha\) are the edges in that layer \(\alpha\) and \(P_\alpha\) is the set of total parameters of the \(X_\alpha\), then between any two nodes in the layer the relation:
\[W_{ij}^{\alpha} = n(X_i^\alpha \cap X_j^\alpha)\]
holds true where, \(W_{ij}^{\alpha}\) is weight of edge \(E_{ij}^\alpha\) , \(i\neq j\) and the parameters of nodes are represented by \(X_i^\alpha , X_j^\alpha \in P_\alpha\).
\par
In layer \(G_\alpha\) we have the graph structure \(G_\alpha=(X_\alpha , E_\alpha)\), where \(X_\alpha\) are the nodes and is given by \(X_\alpha=\{X_1^\alpha,...,X_{N_{\alpha}}^\alpha\}\) and the \(E_\alpha\) are the edges in the layer \(\alpha\). We can calculate the value of similar parameters possessed by any two arbitrary nodes of \(G_\alpha\) from the expression \(n(X_i^\alpha \cap X_j^\alpha)\), where \(1\leq i,j \leq N_\alpha\) and \(1\leq \alpha \leq M\). Now the term \(n(X_i^\alpha \cap X_j^\alpha)\) will be greater if \(X_i^\alpha\) and \(X_j^\alpha\) share more parameters between them. 
We have the total set of parameters \(P_\alpha\)  of layer \(G_\alpha\), then the following relation has to be true:
\[\bigcup_{k=1}^{N_\alpha} X_{k}^\alpha \subseteq P_\alpha\]

\par Now, the adjacency matrix of the layer \(G_\alpha\) can be defined in the following way. The adjacency matrix for layer \(G_\alpha\) is represented as \(A^{[\alpha]}=(a_{ij}^\alpha) \in \Re^{N_\alpha \times N_\alpha}\) given by:\\
\[(a_{ij}^\alpha) =\left\{ \begin{array}{cc} 
                W_{ij}^{\alpha} & \hspace{5mm}   , if (X_i^\alpha , X_j^\alpha) \in E_\alpha \\
                0 & \hspace{5mm}   , otherwise
                \end{array} \right.
\]
where \(A^{[\alpha]}\) is a symmetric matrix,\( 1\leq i,j \leq N_\alpha\) and \(1\leq \alpha \leq M\). 

\begin{equation}
  A^{[\alpha]} = \begin{bmatrix} 
    a_{11} & a_{12} & \dots \\
    \vdots & \ddots & \\
    a_{N_\alpha 1} &        & a_{N_\alpha N_\alpha} 
    \end{bmatrix}
    \label{eq:adj}
\end{equation}

\[
A^{[h]} = \begin{bmatrix} 
    0 & 2 & 1 & 1 \\
    2 & 0 & 1 & 1 \\
    1 & 1 & 0 & 1 \\
    1 & 1 & 1 & 0 \\
    \end{bmatrix}
\]
\[
A^{[d]} = \begin{bmatrix} 
    0 & 1 & 1 & 0 \\
    1 & 0 & 0 & 0 \\
    1 & 0 & 0 & 1 \\
    0 & 0 & 1 & 0 \\
    \end{bmatrix}
;
A^{[p]} = \begin{bmatrix} 
    0 & 1 & 2 & 2 & 1 \\
    1 & 0 & 1 & 1 & 0  \\
    2 & 1 & 0 & 2 & 1  \\
    2 & 1 & 2 & 0 & 1  \\
    1 & 0 & 1 & 1 & 0  \\
    \end{bmatrix}
\]

Where, \(A^{[h]}\), \(A^{[d]}\), \(A^{[p]}\) are the adjacency matrices of the hospital layer, department layer, and doctor layer, respectively, of the network represented in the figure~\ref{fig:toy_network}.
\subsubsection{Inter-layer relationship}In this section, we propose the relational theorem of the inter-layer network. Initially, we propose a general theorem for each layer of the network. We further use a weighted adjacency matrix to capture the overall view of each network.

\paragraph*{Definition 2}

In a layer \(G_{\alpha\beta}=(X_{\alpha\beta},E_{\alpha\beta})\), where \(X_{\alpha\beta}\) are the
nodes of the inter-layer network, \(X_{\alpha\beta} \subseteq X_\alpha \cup X_\beta\),
and \(E_{\alpha\beta}\) are the edges between layers \(\alpha\) and \(\beta\).
Let \(P_\beta\) denote the set of all parameters associated with \(X_\beta\).
Between any two nodes in the layer, the relation is given by
\[
W_{ij}^{\alpha\beta} = n\!\left(P_j^\beta\right),
\]
where \(W_{ij}^{\alpha\beta}\) is the weight of the edge \(E_{ij}^{\alpha\beta}\),
\(P_j^\beta \subseteq P_\beta\), and \(i\neq j\).

The inter-layer adjacency matrix is
\[
A^{[\alpha\beta]} = \big(a_{ij}^{\alpha\beta}\big) \in \mathbb{R}^{N_\alpha \times N_\beta},
\qquad
a_{ij}^{\alpha\beta} =
\begin{cases}
W_{ij}^{\alpha\beta}, & \text{if } (X_i^\alpha, X_j^\beta) \in E_{\alpha\beta},\\[2pt]
0, & \text{otherwise}.
\end{cases}
\]
In general, \(A^{[\alpha\beta]} \neq \big(A^{[\beta\alpha]}\big)^{\mathsf{T}}\),
with \(1 \le i \le N_\alpha\), \(1 \le j \le N_\beta\), and \(1 \le \alpha,\beta \le M\).

\[
A^{[\alpha\beta]} = \begin{bmatrix} 
    a_{11} & a_{12} & \dots a_{1 N_\beta}\\
    \vdots & \ddots & \\
    a_{N_\alpha 1} &        & a_{N_\alpha N_\beta} 
    \end{bmatrix}
\qquad
\]

\[
A^{[hd]}=\begin{bmatrix}
    2 & 1 & 1 & 0 \\
    1 & 2 & 0 & 0 \\
    2 & 0 & 0 & 1 \\
    1 & 0 & 0 & 0 \\
\end{bmatrix}
;
A^{[dp]} = \begin{bmatrix} 
    10 & 6 & 8 & 0 & 0 \\
    0 & 0 & 4 & 6 & 0  \\
    0 & 8 & 0 & 0 & 4  \\
    0 & 0 & 0 & 0 & 6  \\
    \end{bmatrix}
\]
Where \(A^{[hd]}\), \(A^{[dp]}\) are the inter-layer adjacency matrices of the hospital-department layer and department-doctor layer, respectively, of the given network represented in figure~\ref{fig:toy_network}. In our proposed model, we have considered the number of doctors in the department in a hospital as the edge weight of the hospital-department inter-layer network. In contrast, we have considered the qualification of the doctor quantified in terms of number(on a scale of $5$) as the edge weight of the department-doctor inter-layer network.

\subsection{Trust Network }\label{s:trust_net} Trust is central to social interactions within any social system and can be seen as a combination of various incentives that drive these interactions. It encompasses concepts like shared interests, aligned incentives, competence, and knowledge. Therefore, modeling social interactions based on trust relationships among entities in any social system is highly suitable. 
In this section, we define trust in the context of interactions between the medical entities of the proposed network in the healthcare system. We initially categorize the trust into two different types: a) Intra-layer trust and b) inter-layer trust, to explore the two different kinds of interactions in the proposed multi-layered network, i.e., intra-layer and inter-layer interactions. We further define the intra-layer and inter-layer trust matrix to represent the trust network which is derived from the proposed weighted multi-layered network.
\subsubsection{Definition of Trust}We define trust \((\tau_{ij})\) as a perceived importance of an object \(j\) by an another object \(i\). It can be defined as follows:\\

\begin{equation}
    \tau_{ij}=\frac{W_{ij}}{\sum\limits^{\phi}_{k=1}W_{ik}} \label{eq:trust_measure}
\end{equation}

where, \(\phi\) be number of node adjacent to node \(i\). 
\par
\begin{figure}[!htb]
    \includegraphics[width=0.9\textwidth,height=18cm]{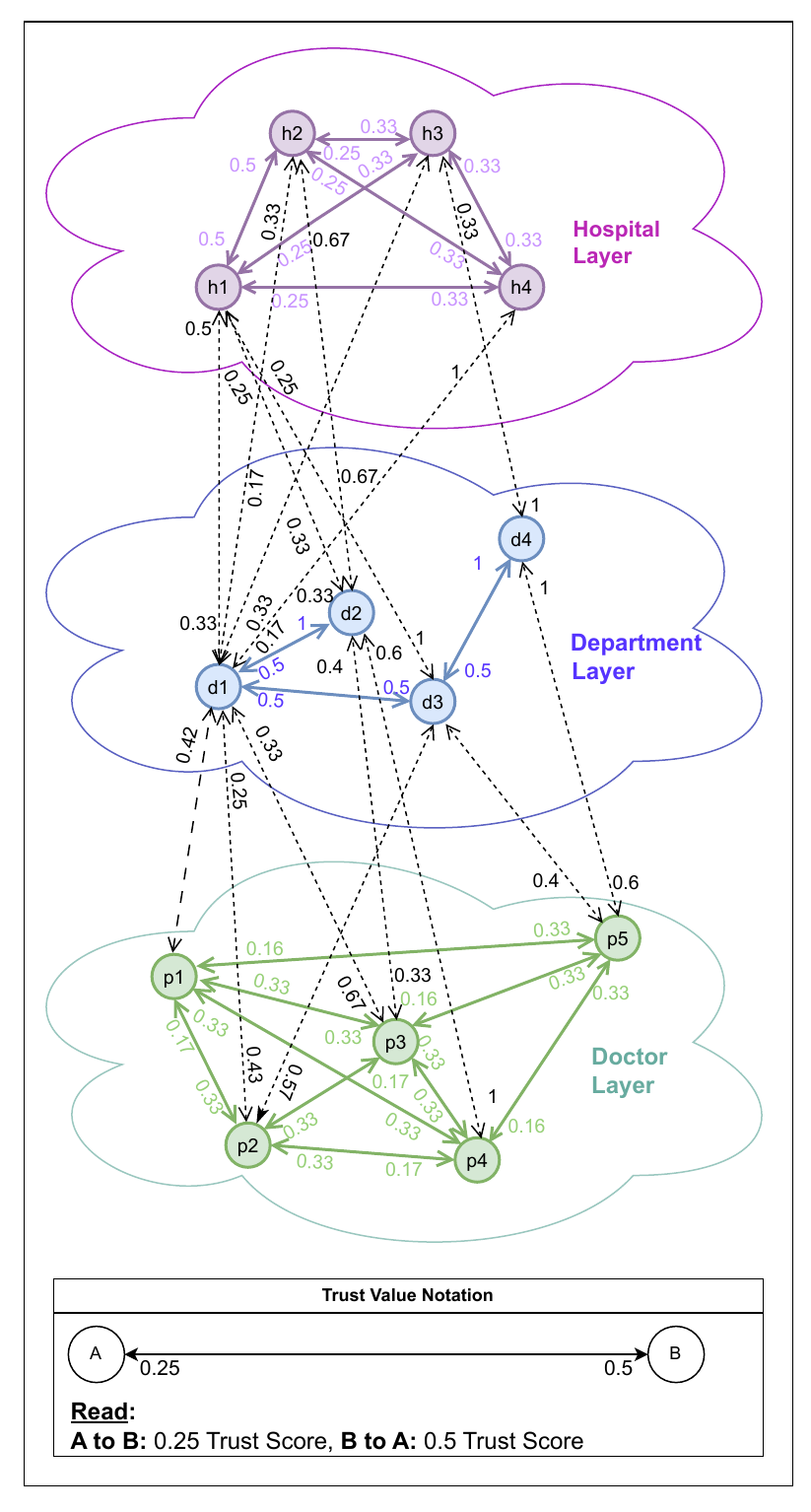}
    \caption{The figure represents the Multi-layer Trust Network with Hospital, Department, and Doctor layers. The nodes and edges of the same color signify they all belong to the same layer. A dotted line represents the inter-layer edge. The edge weight represents the trust value of the directed edge.}
    \label{fig:trust_network}
\end{figure}

 Based on the definition of trust, we build the directed trust network, which is shown in Figure~\ref{fig:trust_network}. The nodes of the network represent the medical entities of the healthcare system, and the trust relationships represent the edges among them. Though the definition of trust is generic across the intra-layer and inter-layer network, the properties of their representative trust matrix are distinct. Therefore, we separately define the trust matrix of intra-layer and inter-layer in the next.
\paragraph{Intra-Layer Trust} 
The intra-layer trust network can be represented as a square non-symmetric matrix. It can be represented as follows:
\begin{equation}
  \tau^{[\alpha]} = \begin{bmatrix} 
    \tau_{11} & \tau_{12} & \dots \\
    \vdots & \ddots & \\
    \tau_{N_\alpha 1} &        & \tau_{N_\alpha N_\alpha} 
    \end{bmatrix} 
    \label{eq:trust_mat}
\end{equation}

where \(\tau^{[\alpha]}\) is a trust matrix of \(\alpha\) layer, \( 1\leq i,j \leq N_\alpha\), \(1\leq \alpha \leq M\) and \( \sum\limits^{\phi}_{k=1}\tau_{ik} = 1 \). Every element of each row \(i\) represents the trust of node \(i\) on other nodes. 

\resizebox{0.47\textwidth}{!}{$
\tau^{[h]} = \begin{bmatrix} 
    0 & 0.5 & 0.25 & 0.25 \\
    0.5 & 0 & 0.25 & 0.25 \\
    0.33 & 0.33 & 0 & 0.33 \\
    0.33 & 0.33 & 0.33 & 0 
    \end{bmatrix} 
\tau^{[d]} = \begin{bmatrix} 
    0 & 0.5 & 0.5 & 0 \\
    1 & 0 & 0 & 0 \\
    0.5 & 0 & 0 & 0.5 \\
    0 & 0 & 1 & 0 
    \end{bmatrix}
    $}
\[
\tau^{[p]} = \begin{bmatrix} 
    0 & 0.17 & 0.33 & 0.33 & 0.16 \\
    0.33 & 0 & 0.33 & 0.33 & 0  \\
    0.33 & 0.17 & 0 & 0.33 & 0.16 \\
    0.33 & 0.17 & 0.33 & 0 & 0.16  \\
    0.33 & 0 & 0.33 & 0.33 & 0  
    \end{bmatrix}
\qquad
\]

\par Here, \(\tau^{[h]}\), \(\tau^{[d]}\), \(\tau^{[p]}\) represent the trust matrix of the hospital layer, department layer, and doctor layer, respectively, of the Figure~\ref{fig:trust_network}.

\paragraph{Inter-Layer Trust}
The inter-layer trust network can be represented as a rectangular matrix. It can be represented as follows:
\[
\tau^{[\alpha\beta]} = \begin{bmatrix} 
    \tau_{11} & \tau_{12}  & \dots \tau_{1 N_\beta}\\
    \vdots & \ddots & \\
    \tau_{N_\alpha 1} &        & \tau_{N_\alpha N_\beta} 
    \end{bmatrix}
\qquad
\]

where \(\tau^{[\alpha\beta]}\) is the trust matrix of the inter-layer trust network for layers \(\alpha\) and \(\beta\), and in general
\(\tau^{[\alpha\beta]} \neq \big(\tau^{[\beta\alpha]}\big)^{\top}\).
Indices satisfy \(1 \le i \le N_\alpha\), \(1 \le j \le N_\beta\), \(1 \le \alpha,\beta \le M\), and the rows are normalized:
\(\sum_{k=1}^{\phi} \tau_{ik} = 1\).

\begin{equation}
\resizebox{0.5\textwidth}{!}{$
\tau^{[hd]} = 
\begin{bmatrix} 
    0.5 & 0.25 & 0.25 & 0 \\
    0.33 & 0.67 & 0 & 0 \\
    0.67 & 0 & 0 & 0.33 \\
    1 & 0 & 0 & 0 
\end{bmatrix}, \quad
\tau^{[dh]} = 
\begin{bmatrix} 
    0.33 & 0.17 & 0.33 & 0.17 \\
    0.33 & 0.67 & 0 & 0 \\
    1.00 & 0 & 0 & 0\\
    0 & 0 & 1.00 & 0 
\end{bmatrix}, 
$}
\end{equation}

\begin{equation}
\resizebox{0.5\textwidth}{!}{$
\tau^{[dp]} = 
\begin{bmatrix} 
    0.42 & 0.25 & 0.33 & 0 & 0 \\
    0 & 0 & 0.4 & 0.6 & 0  \\
    0 & 0.67 & 0 & 0 & 0.33  \\
    0 & 0 & 0 & 0 & 1  
\end{bmatrix}, \quad
\tau^{[pd]} = 
\begin{bmatrix} 
    1.00 & 0 & 0 & 0 \\
    0.43 & 0 & 0.57 & 0  \\
    0.67 & 0.33 & 0 & 0  \\
    0 & 1.00 & 0 & 0  \\
    0 & 0 & 0.40 & 0.60
\end{bmatrix}.
$}
\end{equation}

\par where, \(\tau^{[hd]}\), \(\tau^{[dp]}\) represent the inter-layer trust matrix of the hospital-department layer and department-doctor layer, respectively, of the network given in the Fig~\ref{fig:trust_network}.

\par In the next section, we introduce the concept of social score and discuss the estimation of social score based on trust.
\subsection{Social Score} The social score of any entity of a trust network estimates the trustworthiness of the entity in the network~\cite{oskarsdottir2019value}. It not only captures the trust relations between the nodes in the same layer of the network but also includes the social trust relationships among the different layers. Trust relationships explore the mutual relationship between any two entities in the network. In contrast, social score estimates the importance of the node and helps identify distinguished healthcare entities in the network from various perspectives. 
In this section, we initially define the social score. In this context, we also introduce the concept of residual social score. Further, we estimate the social score based on the residual social score and trust relationship among the entities in the network, as defined in the earlier section.
\par{\bf \em Definition of Social Score:} We can define Social Score \(( S_i^\alpha)\) of an entity \(i\) in a layer \(\alpha\) as the estimated social importance of the entity based on its social interaction in the multi-layer network. More specifically, it could be the cumulative weighted trust of other connected entities of the entity \(i\) in a layer \(\alpha\). It defines how one entity is perceived in the network by other entities.
\par We represent the social score of entities in layer \(\alpha\) as the vector \(\mathbf{S}^\alpha\).
\[
\mathbf{S}^\alpha =
\begin{bmatrix}
s_{1}^\alpha & s_{2}^\alpha & \cdots & s_{N_{\alpha}}^\alpha
\end{bmatrix}.
\]
\par\textbf{\emph{Definition of Residual Social Score:}} Residual social score of any entity \( \delta_i^\alpha \) in a layer \( \alpha \) can be considered as the importance of an entity, which has been based on its feature presumed to be with the entity from its inception.

\par For example, a hospital's residual social score can be estimated based on the location of the hospital, associated brand name, type of the hospital, etc. Similarly, each department has its importance, which can be measured by the number of patients it serves, the importance of organs or body sections dealt with by the department, etc. On the other hand, a doctor's residual social score can be estimated based on his/ her affiliation, work tenure, etc. We represent the residual social score of entities in a layer \(\alpha\) as a vector \((\Delta^\alpha)\).
\[
(\Delta^\alpha) =
\begin{bmatrix}
    \delta_{1} & \delta_{2} & \cdots & \delta_{N_{\alpha}}  
\end{bmatrix}
\]

\par{\bf \em Estimation of Social Score:} In this section, we estimate the social score of the entities of any layer based on the residual social score of the entities and trust matrices of the multi-layered network. We propose the algorithm~\ref{alg:social_score}
to compute the social score of the entities as follows:

\begin{algorithm}
\caption{Estimation of Social Score}\label{alg:social_score}
\begin{algorithmic}
\State \textbf{Input:}$\Delta^\alpha$, $ \Delta^\beta$, $\tau_{\beta\alpha}$
\State \textbf{Output:}$S_{r}^\alpha$
\State \textbf{Initial:} $S_{0}^\alpha \gets \Delta^\alpha + \Delta^\beta * \tau_{\beta\alpha}$
\While{True}
    \State $S_{r+1}^\alpha \gets S_{r}^\alpha * \tau_{\alpha}$\\
    \State $\gamma \gets S_{r+1}^\alpha - S_{r}^\alpha$
    \If{$\gamma \leq 0.001 $ }
        \State $ break $
    \EndIf
\EndWhile
\end{algorithmic}
\end{algorithm}

We can estimate the initial social score\((S_{0}^\alpha)\) as follows:
\[
S_{0}^\alpha = \Delta^\alpha + \Delta^\beta * \tau_{\beta\alpha}
\]
where, layer \(\alpha\) and layer \(\beta\) are adjacent to each other in the multi-layer network {{\textbf{M}}}. We take a feed from the residual social score of layer \(\beta\) to layer \(\alpha\) to estimate the initial social score vector of layer \(\alpha\).

We further solve the recurrence relation, \( S_{r+1}^\alpha \gets S_{r}^\alpha * \tau_{\alpha}\) using the method of substitution to estimate the social score. Thus, further \(S_{r}^\alpha\) can be measured as follows:\(
S_{r}^\alpha = S_{0}^\alpha * (\tau_{\alpha})^{r}
\)

\begin{figure*}[!h]
    \includegraphics[width=\textwidth, height=8cm]{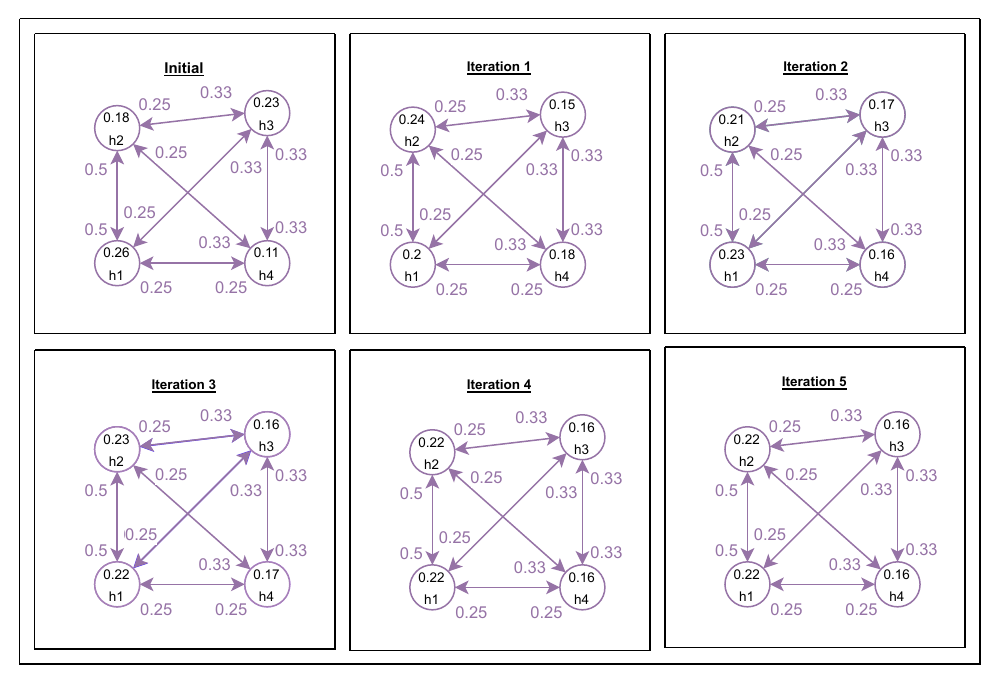}
    \caption{The figure represents iterative social score estimation for hospital-layer nodes in the network (Figure~\ref{fig:toy_network}). The top-left box shows initial hospital social scores; subsequent boxes display score updates per iteration. Residual social scores are uniformly distributed ($\delta_{h}=0.2$, $\delta_{d}=0.2$).}
    \label{fig:social_score}
\end{figure*}

\par{\bf Analysis of Algorithm: }The proposed trust-based algorithm modifies the earlier approach based on Social Trust~\cite{gayen2017towards} by introducing dynamic node interaction weights and an adaptive trust propagation function. In contrast to the static propagation model, we introduce context-aware trust adjustment, where weights are updated based on the frequency and reliability of past interactions, as well as bias-aware filtering, ensuring that skewed trust distributions do not distort evaluation metrics.
\par In the above algorithm, we find that initially, we perform a matrix multiplication of $\Delta^\beta * \tau_{\beta\alpha}$ to estimate the initial social score. In this multiplication involves into two matrices whose order are $1$ x $N_{\beta}$ and $N_{\beta}$ x $N_{\alpha}$. Thus, the time complexity of the initial score calculation would be $O(n)$.
Usually, the Time Complexity of matrix multiplication is $O(n^3)$. However, in the above algorithm, a vector($1$ x $N_{\beta}$) is multiplied with a square matrix$(N_{\alpha}$ x $N_{\alpha})$. Thus, the resulting time complexity is $O(n^2)$. The convergence of social score estimation depends on $\gamma$. Thus, if convergence occurs in the $r$-th iteration, then matrix multiplication occurs $r$ times. Thus, the time complexity of the algorithm is $O(r * n^2)$. For the high-dimensional multi-layered network, we can consider that $r \ll n$. Hence, the overall time complexity of the algorithm is deduced to $O(n^2)$. Though for the $\lim \gamma \to 0$, the time complexity would be reduced to $O(n^3)$, where $r \approx n$. The proposed trust model improves the computational efficiency by optimizing the algorithm’s time complexity from $O(n^3)$ in the work related to social Trust~\cite{gayen2017towards} to $O(n^2)$ in the current model. These enhancements allow for more accurate and fair trust assignments, particularly in complex and dynamic multilayered healthcare systems. We compute the social score of hospital nodes of the network depicted in figure~\ref{fig:toy_network} based on the proposed algorithm~\ref{alg:social_score} and show the convergence of the social score in each iteration in figure~\ref{fig:social_score}. While figure~\ref{fig:social_score} presents a step-by-step example of trust propagation for illustrative purposes on a small subset of nodes, the complete computation of trust scores and social scores in our experiments was performed programmatically on the full dataset. The algorithm iterates until the convergence criterion $\gamma \leq 0.001$ is met, ensuring that all reported results are derived from the full convergence process, not from manually computed values.

\subsection{ Applications of Trust Network: } In this section, we discuss the possible use cases of trust networks and the derived social score of the entities in the healthcare system. The list of use cases is as follows--- 
\begin{itemize}
    \item The social score can be used to develop the recommendation system for various healthcare entities, e.g., Hospitals, Doctors, specific departments of a hospital, etc.
    \item The trust network can be used as an underlying structure to develop a healthcare-based chatbot to interact with external entities, e.g., persons needing help at the time of critical decision-making during the medical emergencies of self or family.  
    \item This type of trust-based healthcare system reduces the risk of self-acclaimed reputations by medical entities. Therefore, it can be used as a robust healthcare network across the country.
   
    \item The trust-based healthcare network can be used as a self-managed extensive database for the Govt. to identify the potential trusted resource for various kinds of needs in the healthcare system, e.g., it could be used for granting funds to any hospital, doctor, or department.   
\end{itemize}
\section{Results \& Discussions}\label{s:result} In this section, we initially describe the dataset, which includes the collection method of the dataset, and its cleaning. We also briefly describe the insight of the dataset. Further, we present the empirical results and discuss the comparative study with the other baseline models. We also outline the deployment challenges of our proposed model to identify possible real-world strategies to counter them at the time of deployment. 

\subsection{Dataset}
In this section, we initially describe the dataset collected in our work. Further, we have developed a graph database to store the collected data, as in our work, relationships among the data play a significant role. We have collected the dataset from Practo~\cite{practo2023}, an online doctor consultation application. We have used web scraping techniques to collect data not only from the Practo web application but also from several hospitals' websites. To store and retrieve the various relational information very quickly from the dataset, we have implemented Neo4j graph database. We have run several queries to evaluate several parameters of the healthcare entities, as well as the weight of the relationship of the proposed multi-layered network.
\subsubsection{Dataset collection Method} 
We used web scraping techniques to collect data from Practo and several hospital websites. In the Practo web application, we can search doctors based on location(city), hospital name, department, etc. Initially, we set a specific area (City) to list down the doctors. Based on the town, it provides the list of doctors who are available on Practo. We collected profiles of doctors that contain the following information regarding the doctors: name, associated hospital/clinic name, qualification, experience (overall experience as well as experience in specialized domain), department(s) associated with, percentage of likes received from the verified patients measured by the number of up and down votes received. We recursively use this method to collect doctor information from several hospitals and their various departments. The distribution of ` like percentage' of the doctors is observed strictly skewed in nature, possibly reflecting bias in the percentage of likes and indicating a risk of manipulative reputation creation. Thereafter, based on the listed doctors in our dataset, we collect detailed information regarding the hospital, e.g., name, rating, and the number of stories of the hospital. We also collect information regarding the departments in which the listed doctors are associated. Though we are not able to find the rating or review of the department, we further compute that based on the other information collected related to the department, e.g., the number of reviews received by the doctors in the department. We also collect the hospital address and accreditation by visiting the corresponding hospital's official website. We also categorise the location (Urban/suburban/rural) of the hospital based on the address collected from the websites\footnote{All code used in this study—including web-scraping scripts, data cleaning/normalization utilities, Neo4j graph builders, and analysis/visualization pipelines—is openly available at: $https://github.com/Samsomyajit/Trust_Health$}. 
\newcommand{\head}[1]{\textbf{#1}}

\begin{table}[h!]
\centering
\begin{tabular}{||p{0.029\textwidth}|p{0.075\textwidth}|p{0.029\textwidth}|p{0.075\textwidth}|p{0.029\textwidth}|p{0.075\textwidth}||} 
 \hline
 \multicolumn{2}{||c|}{\head{Doctor}} &
  \multicolumn{2}{c|}{\head{Hospital}}& 
   \multicolumn{2}{c||}{\head{Department}}\\
 \hline
 Raw & Filtered & Raw & Filtered & Raw & Filtered\\
 \hline\hline
 160 & 77 & 15 & 10 & 53 & 32\\
 \hline
\end{tabular}
\caption{General Information of Dataset.}
\label{table:dataset_stat}
\end{table}

\begin{table}[ht]
\centering
\caption{Doctor Specialization Distribution Information}
\label{tab:Doct_dataset_SpeDistri}
\renewcommand{\arraystretch}{1.2} 
\setlength{\tabcolsep}{4pt} 
\begin{tabular}{|p{0.15\columnwidth}|p{0.45\columnwidth}|p{0.2\columnwidth}|}
\hline
\textbf{Sl. No.} & \textbf{Specialization} & \textbf{Doctor Count} \\ \hline
1 & Gynecologist & 69 \\ \hline
2 & Therapist & 5 \\ \hline
3 & Laparoscopic Surgeon & 9 \\ \hline
4 & Orthopedic Surgeon & 10 \\ \hline
5 & Spine and Pain Specialist & 5 \\ \hline
6 & Spine Surgeon & 3 \\ \hline
7 & Pulmonologist & 6 \\ \hline
8 & General Physician & 19 \\ \hline
9 & Head and Neck Surgeon & 3 \\ \hline
10 & ENT/Otorhinolaryngologist & 7 \\ \hline
11 & Plastic Surgeon & 3 \\ \hline
12 & Diabetic Foot Surgeon & 2 \\ \hline
13 & Hair Transplant Surgeon & 4 \\ \hline
14 & Burn Surgeon & 3 \\ \hline
15 & Psychiatrist & 6 \\ \hline
16 & Addiction Psychiatrist & 2 \\ \hline
17 & Neuropsychiatrist & 2 \\ \hline
18 & Geriatric Psychiatrist & 2 \\ \hline
19 & Clinical Psychologist & 2 \\ \hline
20 & Sexologist & 2 \\ \hline
\end{tabular}
\end{table}

\subsubsection{Data cleaning} Finally, we have removed those doctors' profiles that are `unclaimed'  or `not-verified' from our listed raw dataset. We have further removed those doctors' profiles that were missing a few details. Further, based on the filtered dataset of doctors, we have collected the hospital data and the department data of those doctors. We removed those hospitals/ clinics that are unrated. In table~\ref{table:dataset_stat}, we show the general statistics of our collected dataset. In the table~\ref{tab:Doct_dataset_SpeDistri}, we show the doctor specialization distribution of our collected dataset.

\subsection{Empirical Results}
In this section, we observe the results of two experiments performed in our work---a) distribution of trust values of the three different layers, b) distribution of social score of the nodes in the three different layers of the multi-layered healthcare network. The distribution of trust values across the different layers would disclose the nature of the mutual trust relationship among the different nodes in each layer, and would also identify the variation of trust relationships across the different layers in the healthcare system. More specifically, from this observation, we will be able to understand the possible framework of patient referral among doctors and hospitals. It would also find the strong need for the establishment of various departments together in a hospital to provide emergency services. The referral among the hospitals and strong trust relations may reveal the need for collaborative activities among them.
On the other hand, the distribution of social scores across the different layers would reveal the variation in the social importance of the healthcare entities. It would also help us compare and validate the proposed social score values based on the rating of each entity. Understanding the distribution of social scores would help us to develop a bias-free recommendation system.

\subsubsection{Distribution of Trust in different layers} 
\begin{figure}[htb]
    \centering
    \includegraphics[width=0.8\textwidth]{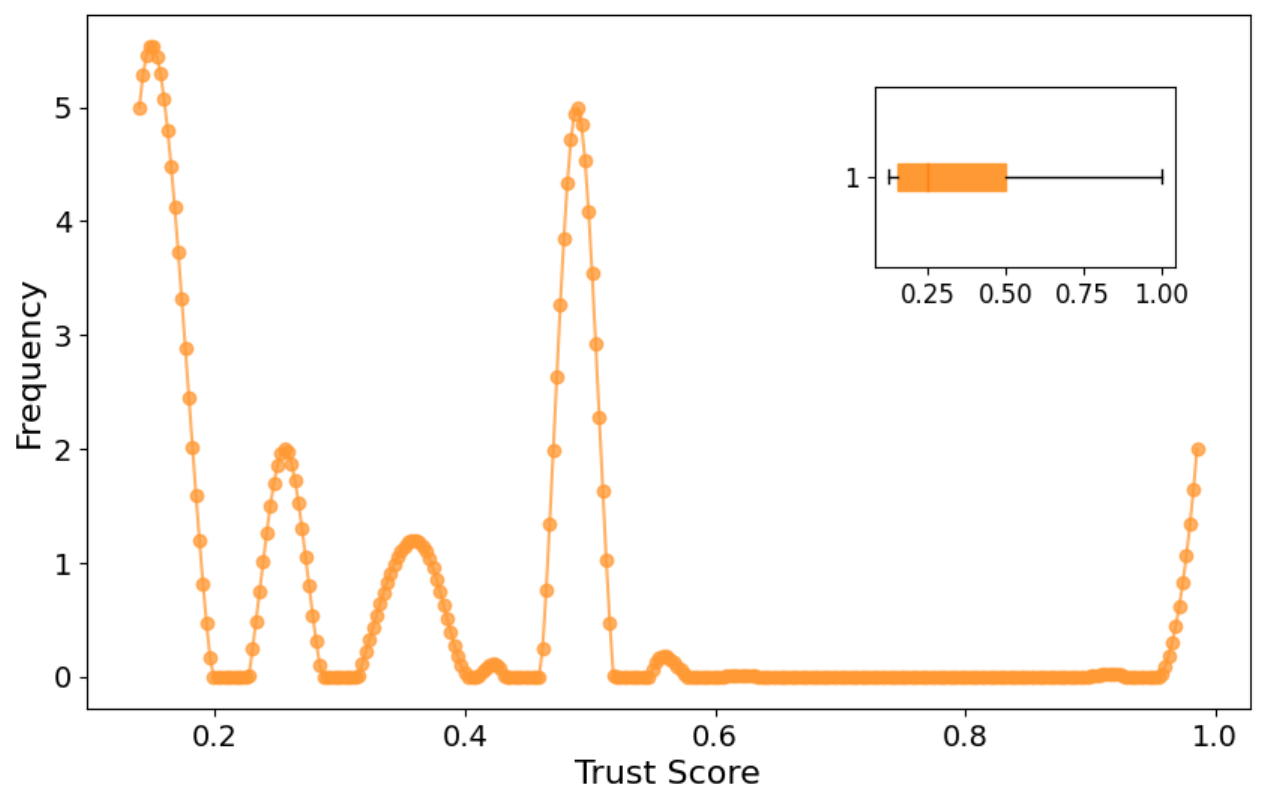}
    \caption{Distribution of non-zero trust score among the hospitals. Along the x-axis, we plot the trust score, which ranges between 0.0 and 1.0, and along the y-axis, it represents the frequency. In the inset, we plot the same data in a box plot to represent the min, first quartile, median, third quartile, and max.
    }
    \label{fig:trust_score_dis_Hos}
\end{figure}

\begin{figure}[htb]
    \centering
    \includegraphics[width=0.8\textwidth]{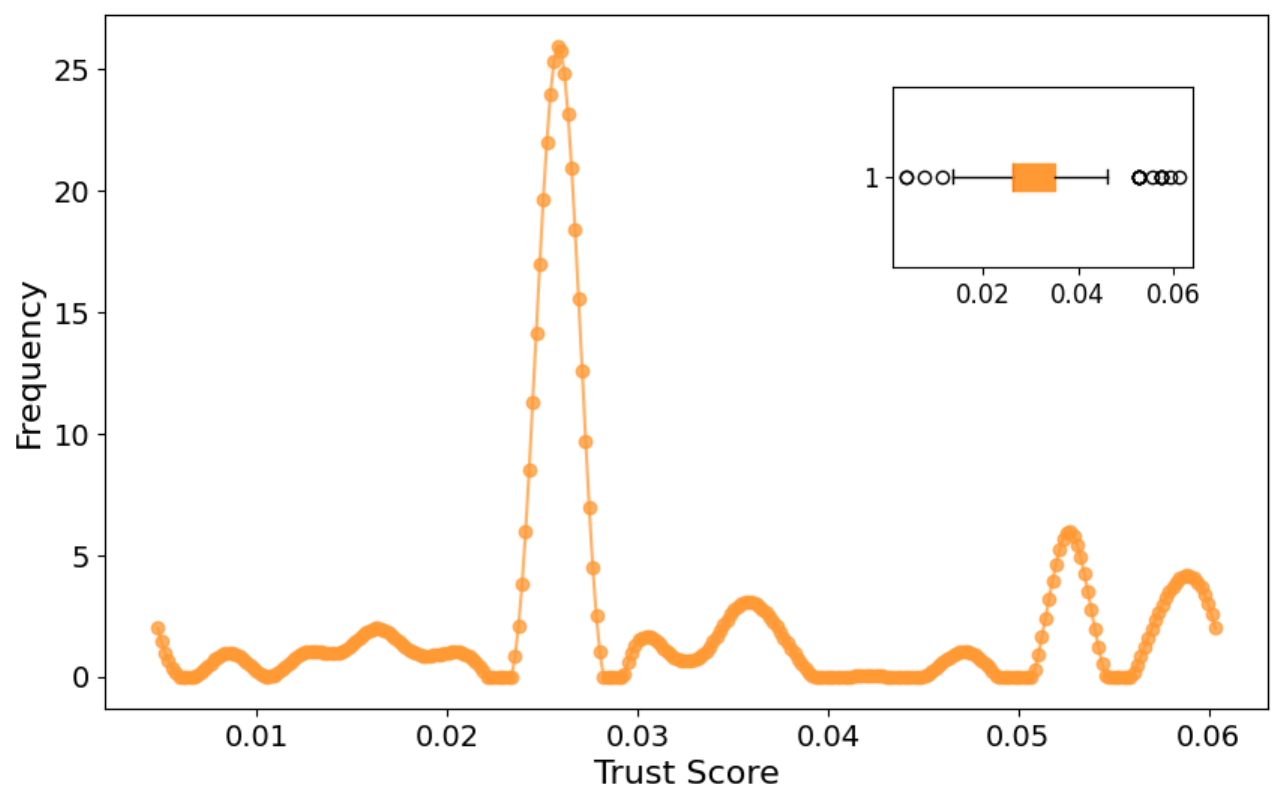}
    \caption{Distribution of non-zero trust score among the departments. Along the x-axis, we plot the trust score, which ranges between 0.0 and 1.0, and along the y-axis, it represents the frequency. In the inset, we plot the same data in a box plot to represent the min, first quartile, median, third quartile, and max.
    }
    \label{fig:trust_score_dis_Dept}
\end{figure}

\begin{figure}[htb]
    \centering
    \includegraphics[width=0.8\textwidth]{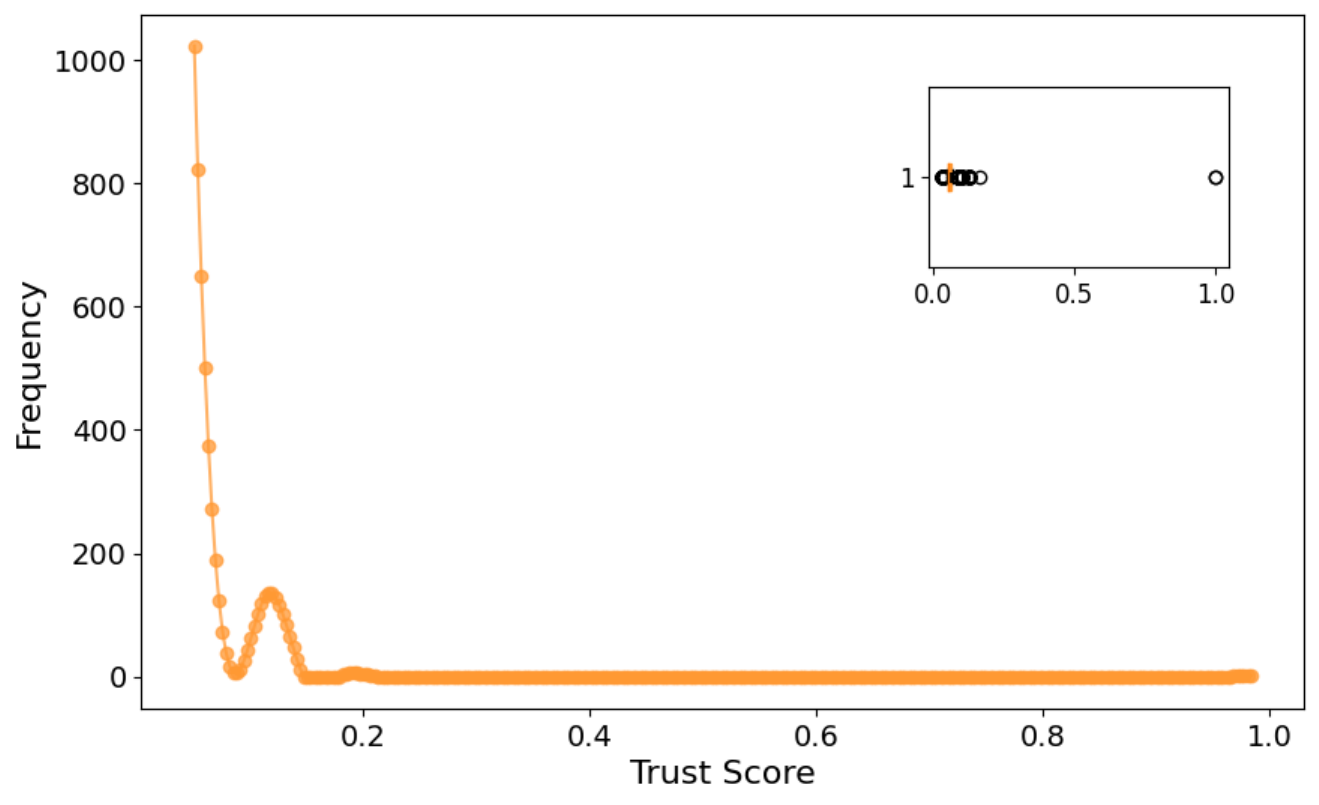}
    \caption{Distribution of non-zero trust score among the doctors. Along the x-axis, we plot the trust score ranges between 0.0 and 1.0, and along the y-axis, it represents the frequency. In the inset, we plot the same data in a box plot to represent the min, first quartile, median, third quartile, and max. }
    \label{fig:trust_score_dis_Doc}
\end{figure}

Though the inter-layer trust relationship captures the impact of one layer on the other to estimate the social score of the nodes in each layer, we only observe the intra-layer trust relationship among the nodes, as understanding the nature of intra-layer would help us to understand the intricate relationship among the healthcare entities. Initially, we develop the multi-layered healthcare network and then, based on the multi-layered network, we create the trust network using the crawled data. We compute the trust score of each edge among the nodes in each layer. In this section, we observe the trust value distribution of each layer of the network separately. In figure~\ref{fig:trust_score_dis_Hos}, ~\ref{fig:trust_score_dis_Dept} and \ref{fig:trust_score_dis_Doc}, we represent the distribution of the trust values of the edges of the hospital, department, and doctor layers, respectively. The trust score of each edge of the layer is estimated based on the equation~\ref {eq:trust_measure}. We evaluate the trust score on values given in the adjacency matrices (eq.~\ref{eq:adj}) to obtain trust matrices (eq.~\ref{eq:trust_mat}), where each element represents the trust score between entities. Though the trust matrix is an n × n matrix, the total number of trust scores corresponding to each trust matrix is 100, 1024, and 5929 for hospitals, departments, and doctors, respectively. As the network of each layer does not form a complete graph, there exist many trust scores that are zero. Therefore, for more precise observation, we exclude those trust scores. Finally, we observe $20$ non-zero trust values for the hospital layer, $64$ for departments, and $1172$ for doctors. This adjustment provided a more precise representation of trust score distributions, highlighting meaningful trust relationships among hospitals, departments, and doctors. The figures now accurately depict the frequency distribution of non-zero trust scores, enhancing the interpretability of the data.

\par Figure \ref{fig:trust_score_dis_Hos} shows the distribution of trust score of the mutual relationship among the hospitals. It shows that the distribution is highly skewed towards the lower end. It shows that though the trust score varies a wide range (0-1), there are many mutual relations amongst the hospitals that are low in trust score (0-0.2), whereas there are comparatively fewer relations that have a high trust score. The majority of hospitals have low to moderate trust scores, with a notable concentration around the lower end of the spectrum. The inset box plot reinforces this observation, indicating that most trust scores are clustered towards the lower quartile, with a few outliers extending into higher trust scores. It shows the power-law curve, which corroborates the similar nature in other major social networks. 

\par In contrast, Figure 15, which displays the distribution of trust scores in the department layer, reveals a more detailed distribution with frequent peaks and troughs, particularly around 0.03. This suggests variability and possibly a mixture of departments with different trust levels. The department trust score distribution is more evenly spread compared to hospitals, with several peaks and valleys in the smooth line chart. The box plot inset indicates a broader interquartile range, highlighting a more diverse trust score distribution among departments.
\par In figure~\ref{fig:trust_score_dis_Doc}, we observe that the trust score distribution among doctors, showing a wide range of trust scores (0-1), with the majority of mutual relations having low trust scores (0-0.2). Its nature is similar to the trust score distribution of the hospital and department layers. Figure~\ref{fig:trust_score_dis_Doc} shows that the trust score is skewed towards lower trust scores, with a few peaks indicating higher frequencies at specific intervals. The smooth line chart, complemented by the box plot, shows a concentration of scores in the lower quartile and a few outliers with higher trust scores. We exclude zero or near-zero trust scores for clear visualizations of the trust score distributions. It highlights meaningful trust relationships among the doctors. 
\subsection{Quantitative Benchmarking \& Baselines}
To validate the effectiveness of our trust score calculations, we benchmark our proposed model against existing trust-based models in the literature. The comparison, as shown in Table~\ref{tab:benchmark}, highlights the superior accuracy and nuanced inter-entity dynamics captured by our approach. Additionally, we conduct quantitative benchmarking and compare it with baseline models to demonstrate the effectiveness of our proposed approach. Finally, we include a section featuring stress testing using synthetic data, which validates the generalizability of our model beyond the limited scraped dataset. 

\begin{table}[ht]
\centering
\caption{Benchmarking Trust Score Distributions}
\label{tab:benchmark}
\begin{tabularx}{\columnwidth}{|X|X|X|X|}
\hline
\textbf{Metric} & \textbf{Proposed Model} & \textbf{Guo et al. \cite{guo2016doctor}} & \textbf{Mondal et al. \cite{mondal2020building}} \\ \hline
\textbf{Trust Score Range} & 
Skewed (most 0.0–0.2) with a long tail & 
Uniform & 
Skewed lower-range \\ \hline

\textbf{Accuracy} & 
Correlation $\approx$ 0.91 for departments and hospitals; lower for doctors & 
Not specified & 
Correlation $\approx$ 0.75 \\ \hline

\textbf{Inter-Entity Dynamics} & 
Captures the interplay among hospitals, doctors, and departments & 
Limited to doctor-patient links & 
No inter-layer dynamics \\ \hline

\textbf{Scalability} & 
Efficient ($O(\cdot n^2)$) for medium datasets & 
Not analyzed & 
Not analyzed \\ \hline

\textbf{Context Awareness} & 
Uses hospital ratings, doctor qualifications, and department importance & 
Limited & 
Patient-centric \\ \hline

\textbf{Bias Mitigation} & 
Reduces rating biases (e.g., outliers) & 
No mitigation & 
Limited \\ \hline

\end{tabularx}
\end{table}

\subsubsection{Distribution of Social Score in different layers} In this section, we observe the distribution of the social score of the healthcare entity. We compute the social score based on the trust network developed using the collected dataset. We execute the algorithm~\ref{alg:social_score} to estimate the social score of healthcare entities. We consider three different kinds of residual social score distribution, e.g., uniform distribution, normal distribution, and skewed distribution, of the healthcare entities to understand the possible real-life scenarios of the development of such a network. We further estimate the correlation coefficient of the social score with the rating value of the entities to validate the correctness of the social score in the healthcare system. Although we have collected the like \% of the doctors using the web crawling, further we have mapped the like \% in rating score ranges between $0-5$, which we have used for the validation of social score of the doctors.
 
%

\par 
\par Initially, the proposed algorithm defines the residual social score that each entity has by its intrinsic features, such as location, brand, or departmental importance. The base social score computation takes these residual scores and combines them with trust matrices representing relationships and interactions at each layer of the multi-layered network. These scores slowly come together with every iteration made by the algorithm and, in the end, minimize variance between every pair of successive iterations. This is very efficient in terms of computational complexity, which scales quadratically with the number of entities and proves to be highly beneficial in high-dimensional networks. It can usually be approximated under different simulation scenarios for various types of distributions—Uniform, Normal, Skewed—to provide quite an excellent framework for checking the influence of data distribution on social scores. 
\par
\begin{figure*}[!hbt]
    \includegraphics[width=1\textwidth]{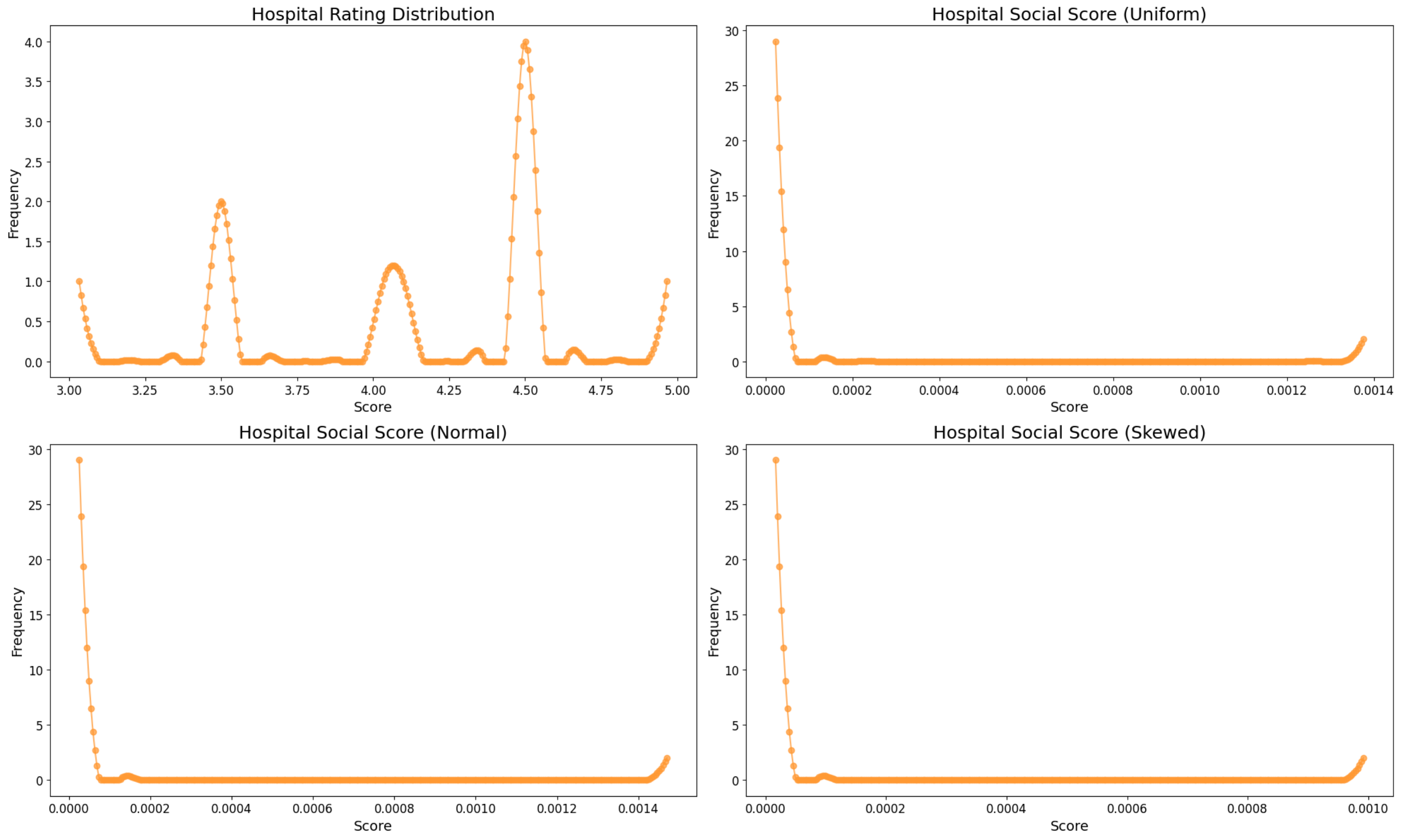}
    \caption{The figure represents the distribution of ratings (top left) and social scores based on a) Uniform (top right), b) Normal (bottom left), and c) Skewed (bottom right) residual social scores for the Hospital layer}
    \label{fig:hospital_social_score_distribution}
\end{figure*}

\begin{figure*}[!hbt]
    \includegraphics[width=1\textwidth]{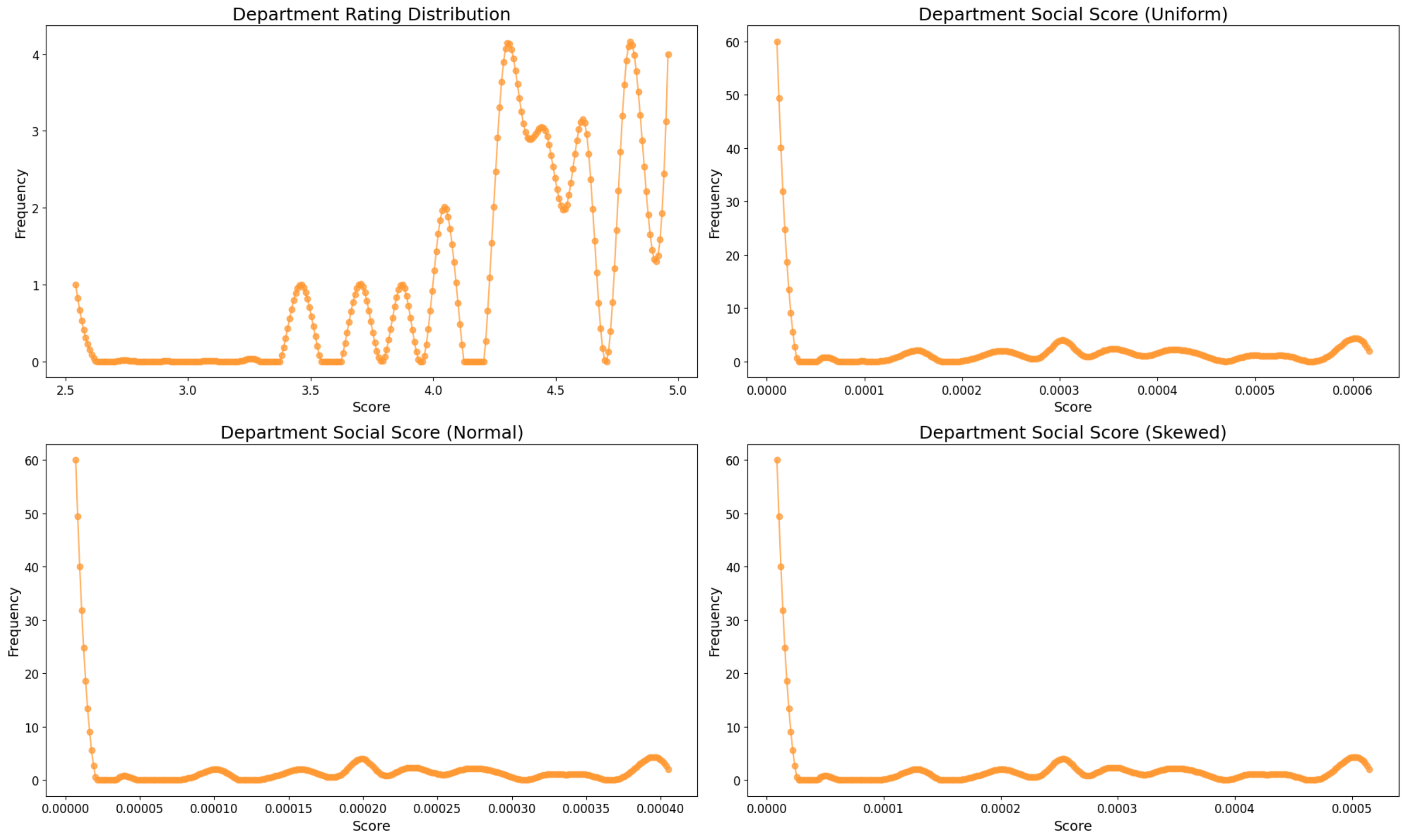}
    \caption{The figure represents the distribution of ratings (top left) and social scores based on a) Uniform (top right), b) Normal (bottom left), and c) Skewed (bottom right) residual social scores for the Department layer}
    \label{fig:department_social_score_distribution}
\end{figure*}

\begin{figure*}[!hbt]
    \includegraphics[width=1\textwidth]{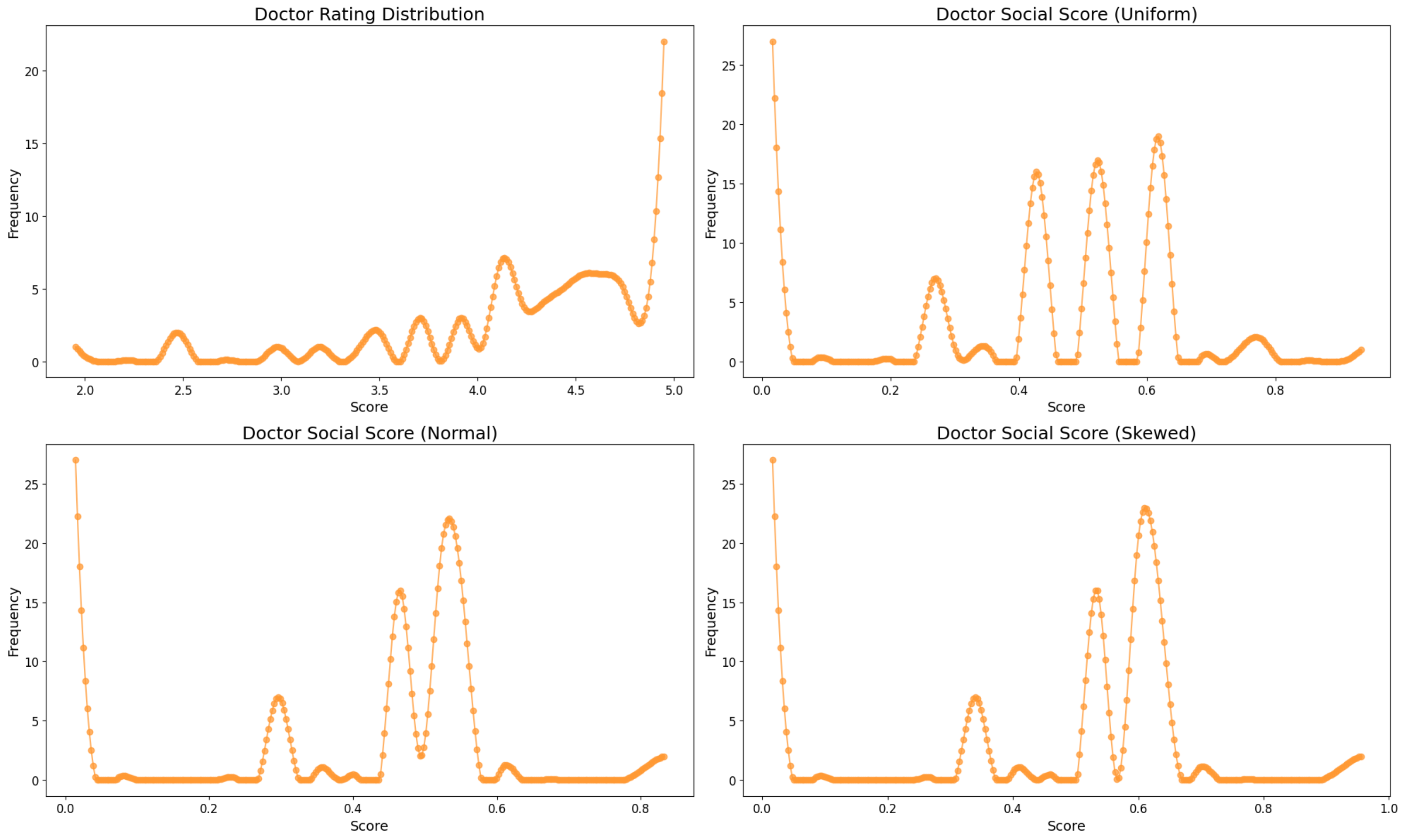}
    \caption{The figure represents the distribution of ratings (top left) and social scores based on a) Uniform (top right), b) Normal (bottom left), c) Skewed (bottom right) residual social scores for the Doctor layer}
    \label{fig:doctor_social_score_distribution}
\end{figure*}

 Figure \ref{fig:hospital_social_score_distribution},\ref{fig:department_social_score_distribution}, and \ref{fig:doctor_social_score_distribution} represent the distribution of social scores generated by the proposed algorithm of the three different layers: hospital, department, and doctor, along with the distribution of ratings of the entities in that layer. From these observations, we can infer that the initial distribution of residual social scores significantly impacts the resulting social score distributions of the various layers. The uniform distribution scenario tends to smooth out variations, leading to a more consistent pattern across all layers. In contrast, the skewed distribution emphasizes lower social scores more prominently, particularly in the department layer, suggesting that initial biases in residual scores can amplify certain trust relationships. The typical distribution scenario provides a middle ground, balancing between the extremes observed in the other two scenarios. This variability indicates that the model's sensitivity to initial conditions must be carefully considered when interpreting trust scores.

\begin{table}[ht]
\centering
\begin{tabular}{lccc}
\hline
\textbf{Layer}  & \textbf{Uniform} & \textbf{Normal} & \textbf{Skewed} \\ \hline
Hospital       & -0.0015          & -0.0015         & -0.0015         \\
Department     & 0.9088           & 0.9088          & 0.9088          \\
Doctor         & 0.4362           & 0.4667          & 0.4627          \\ \hline
\end{tabular}
\caption{Spearman Correlation Scores for Ratings and Social Scores}
\label{tab:spearman_correlation}
\end{table}

\par 
The analysis of social score distributions across hospital, doctor, and department layers reveals several insights into social scores and ratings. Figures \ref{fig:hospital_social_score_distribution}, \ref{fig:department_social_score_distribution} and \ref{fig:doctor_social_score_distribution} display distinct patterns for each layer. For the hospital layer results shown in Figure \ref{fig:hospital_social_score_distribution}, the rating distribution shows peaks around 4.25 to 4.75, with a frequency drop-off at higher scores. In contrast, the social scores for hospitals are mostly concentrated at very low values, indicating a right-skewed distribution. Similarly, in Figure \ref{fig:department_social_score_distribution}, department ratings cluster around 4.0 to 4.5, with lower social scores predominantly concentrated near zero, suggesting departments have high ratings but low variability in social scores. In Figure \ref{fig:doctor_social_score_distribution}, doctors exhibit a more diverse range of scores, with notable peaks in ratings around 4.0 to 5.0 and varying social scores, suggesting a wider distribution of trust and performance metrics.
\par Table \ref{tab:spearman_correlation} highlights the Spearman correlation scores between ratings and social scores across different layers, with hospitals showing a correlation of -0.0015, indicating a negligible relationship between ratings and social scores. Departments correlate 0.9088, reflecting a strong positive relationship, implying that higher department ratings are closely aligned with higher social scores. Doctors have correlations ranging from 0.4362 to 0.4667, indicating a moderate positive relationship, suggesting that doctor ratings moderately correlate with their social scores.
\par These findings underscore the varying nature of ratings across different layers. Hospitals, despite high ratings, show little correlation with social scores, possibly due to factors external to trust influencing the ratings. Departments exhibit strong alignment between ratings and social scores, suggesting internal consistency and reliability in performance assessments. Doctors show moderate correlation, indicating that while ratings and social scores are related, other factors may also play significant roles in their evaluations.

\subsubsection{Quantitative Benchmarking \& Baselines}
To assess the practical effectiveness of our model, we computed multiple evaluation metrics (Precision@k, Recall@k, F1, RMSE, MAE, Spearman, Kendall) for each layer of the healthcare system (Hospital, Department, Doctor). As shown in the results, Hospitals and Departments generally exhibit high precision and recall values at lower k-values, with high Spearman and Kendall correlations, particularly for the hospital and department baselines. In contrast, Doctors show a decrease in precision and recall, indicating more variability in social-score ratings for individual practitioners. The RMSE and MAE values are computed for each baseline (e.g., Doctor\_Count, Stories\_Count, Accreditation\_Score) against the ground-truth ratings. Notably, Doctor-level metrics exhibit higher RMSE/MAE, signaling greater deviation in the inferred social scores compared to the ratings.

\begin{table}[t]
\centering
\caption{Post-hoc Quantitative Benchmarking for Hospitals, Departments, and Doctors}
\label{tab:quant_benchmark}
\setlength{\tabcolsep}{3.5pt}       
\renewcommand{\arraystretch}{1.1}   
\resizebox{\columnwidth}{!}{%
\begin{tabular}{l l c c c c c c c c c}
\toprule
\textbf{Layer} & \textbf{Baseline} & \textbf{K} & \textbf{Precision} & \textbf{Recall} & \textbf{F1} & \textbf{RMSE} & \textbf{MAE} & \textbf{Spearman} & \textbf{Kendall} & \textbf{N} \\
\midrule
\multirow{4}{*}{Hospital} & Doct\_Count & @3 & 1.000 & 1.000 & 1.000 & NaN & NaN & NaN & NaN & 3 \\
 & Doct\_Count & @5 & 0.600 & 0.600 & 0.600 & NaN & NaN & NaN & NaN & 3 \\
 & Dept\_Count & @3 & 1.000 & 1.000 & 1.000 & NaN & NaN & NaN & NaN & 3 \\
 & Dept\_Count & @5 & 0.600 & 0.600 & 0.600 & NaN & NaN & NaN & NaN & 3 \\
\midrule
\multirow{4}{*}{Department} & Doctor\_Count & @5 & 0.600 & 0.600 & 0.600 & 0.000000 & 0.000000 & 1.000 & 1.000 & 3 \\
 & Doctor\_Count & @8 & 0.375 & 0.375 & 0.375 & 0.000000 & 0.000000 & 1.000 & 1.000 & 3 \\
 & Doct\_Review\_Count & @5 & 0.600 & 0.600 & 0.600 & 0.032075 & 0.018519 & 1.000 & 1.000 & 3 \\
 & Doct\_Review\_Count & @8 & 0.375 & 0.375 & 0.375 & 0.032075 & 0.018519 & 1.000 & 1.000 & 3 \\
\midrule
\multirow{4}{*}{Doctor} & Vote\_Count & @10 & 0.300 & 0.300 & 0.300 & 0.609483 & 0.497635 & 0.500 & 0.333333 & 3 \\
 & Vote\_Count & @20 & 0.150 & 0.150 & 0.150 & 0.609483 & 0.497635 & 0.500 & 0.333333 & 3 \\
 & Review\_Count & @10 & 0.300 & 0.300 & 0.300 & 0.831205 & 0.821429 & -0.500 & -0.333333 & 3 \\
 & Review\_Count & @20 & 0.150 & 0.150 & 0.150 & 0.831205 & 0.821429 & -0.500 & -0.333333 & 3 \\
\midrule
\multirow{4}{*}{Doctor} & Ovr\_Exp & @10 & 0.300 & 0.300 & 0.300 & 0.755501 & 0.712366 & -0.500 & -0.333333 & 3 \\
 & Ovr\_Exp & @20 & 0.150 & 0.150 & 0.150 & 0.755501 & 0.712366 & -0.500 & -0.333333 & 3 \\
 & Yr\_As\_Spe & @10 & 0.300 & 0.300 & 0.300 & 0.748645 & 0.698276 & -0.500 & -0.333333 & 3 \\
 & Yr\_As\_Spe & @20 & 0.150 & 0.150 & 0.150 & 0.748645 & 0.698276 & -0.500 & -0.333333 & 3 \\
\bottomrule
\end{tabular}%
}
\end{table}

\par The results of the quantitative benchmarking presented in Table~\ref{tab:quant_benchmark} show the performance of the trust-based social score model across three layers: Hospital, Department, and Doctor. For the Hospital layer, the baseline metrics such as Doctor Count (Doct\_Count) and Department Count (Dept\_Count) show excellent performance at lower k-values (e.g., Precision@3 and Recall@3), achieving a perfect 1.000 for both precision and recall in most cases. This indicates that the top-ranked hospitals, according to the number of doctors or departments, align perfectly with the ground-truth ratings. However, as we move to higher k-values (e.g., @5), precision and recall decrease to 0.600, suggesting that as we consider more hospitals, the model’s performance weakens, this behavior is consistent across both Doct\_Count and Dept\_Count, showing that while the baseline features help rank the top hospitals, they are less effective when a broader set of hospitals is considered. Notably, NaN values are observed in the RMSE and MAE columns for the Hospital layer, which likely indicates that these metrics were not computed for these baselines, as the trust score model for hospitals did not produce valid social score predictions for these entities.

\par For the Department layer, the baselines Doctor Count (Doctor\_Count) and Doctor Review Count (Doct\_Review\_Count) demonstrate a similar trend to the hospital layer. At lower k-values (e.g., @5), the model achieves a precision and recall of 0.600, but these values drop when considering more departments (e.g., @8), with precision and recall both dropping to 0.375. This suggests that while the size of a department (measured by doctor count) and the level of engagement (measured by doctor reviews) are useful for ranking departments, they become less reliable as the number of departments increases. The RMSE and MAE for Doct\_Review\_Count are relatively low, indicating a good alignment between the predicted social scores and the actual ratings for department-level scores. Moving to the Doctor layer, the performance metrics drop considerably. Vote Count (Vote\_Count) and Review Count (Review\_Count) show much lower precision and recall values at both @10 and @20, indicating that these baselines do not effectively predict the ratings of doctors, especially when considering more doctors. The RMSE and MAE for doctors are significantly higher than those for hospitals and departments, pointing to a higher deviation between the inferred social scores and the ground-truth ratings for individual practitioners. Additionally, the Spearman and Kendall correlations for doctors show weaker alignment with ground-truth ratings compared to hospitals and departments, with values as low as -0.5 for some baselines. The NaN values for the RMSE and MAE metrics in specific rows indicate that for these particular baselines, the model was unable to compute valid predictions, likely due to the complexity or sparsity of the data for individual doctors.

\subsubsection{Synthetic Trust-Graph Stress Test (CTGAN)}To assess generalizability beyond our small scraped dataset (77 doctors, 32 departments, 10 hospitals), we generated \emph{synthetic} trust matrices using a tabular Generative Adversarial Network (CTGAN). We trained a single-table conditional generator on the edge-level table $\{(\text{layer}, \text{src}, \text{dst}, \text{trust})\}$ constructed from the paper’s trust matrices ($\tau^{[h]}, \tau^{[d]}, \tau^{[p]}, \tau^{[hd]}, \tau^{[dh]}, \tau^{[dp]}, \tau^{[pd]}$). After sampling, we rebuilt each $\hat{\tau}$ by pivoting to matrix form and enforcing row-stochasticity; social scores were then recomputed with the same damped propagation used in the main model. This follows standard practice for synthetic tabular data generation and evaluation (e.g., CTGAN/SDV~\cite{espinosa2023quality},\cite{yadav2024rigorous}), while preserving our modeling constraints (row-stochastic $\tau$, zero diagonals in intra-layer matrices).

\par Across layers, the synthetic $\hat{\tau}$ preserves rankings of the trust-driven social scores $S$ well for Hospitals and Doctors. Using the true $S$ as reference and the synthetic $\hat{S}$ as predictions, we obtain: \emph{Hospital} $P@3=P@4=1.00$, $R@3=R@4=1.00$, $F1=1.00$, $\text{Spearman}=1.0$, $\text{Kendall}=1.0$, $\text{RMSE}=0.445$, $\text{MAE}=0.273$; \emph{Doctor} $P@3=0.667$, $P@5=1.00$ ($R@k=F1@k$ identically), $\text{Spearman}=0.9$, $\text{Kendall}=0.8$, $\text{RMSE}=0.464$, $\text{MAE}=0.309$. The \emph{Department} layer shows weaker rank preservation ($P@3=0.667$, $P@4=1.00$, $\text{Spearman}=-0.4$, $\text{Kendall}=-0.333$, $\text{RMSE}=0.666$, $\text{MAE}=0.573$) due to the tiny $4{\times}4$ matrix, heavy zero mass, and ties that amplify small perturbations as explored in the diagnostic tests.

\par
\begin{figure*}[htbp]
    \centering
    \includegraphics[width=1\textwidth, height = 3cm]{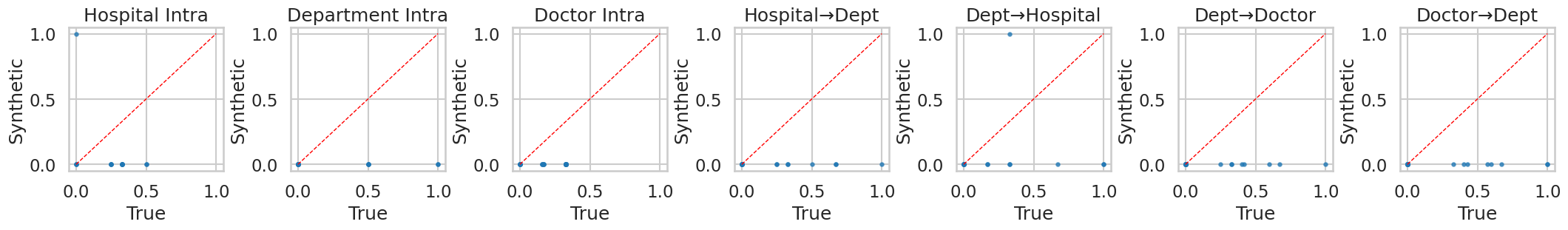}
    \caption{Scatter plots of true vs. synthetic trust values ($\tau_{ij}$) for each intra- and inter-layer trust matrix.
    The horizontal axis shows the trust score from the true dataset, and the vertical axis shows the trust score from the synthetic (CTGAN-generated) dataset.
    Each blue point corresponds to one directed edge $(i,j)$.
    The red dashed line indicates the ideal $y = x$ diagonal; points on or near this line indicate perfect agreement between true and synthetic trust scores.}
    \label{fig:syn_vs_true}
\end{figure*}

\begin{figure}[htbp]
    \centering
    \includegraphics[width = 0.8\textwidth]{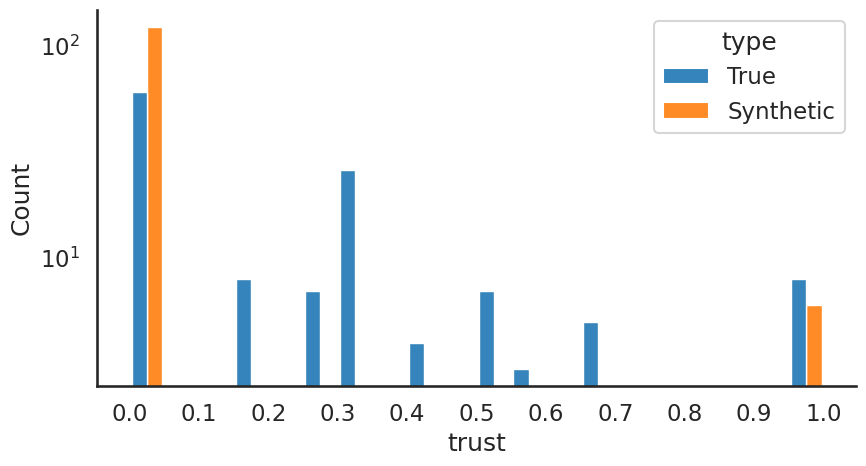}
    \caption{Distribution of trust scores (in log scale) in the true dataset (blue) and synthetic dataset (orange), aggregated across all layers.
    The horizontal axis shows the trust value ($\tau_{ij}$) and the vertical axis shows the number of edges (count) with that trust value. The synthetic data matches the overall zero-mass concentration of the true data, but exhibits fewer medium-strength trust edges in small, sparse layers such as the Department layer.}
    \label{fig:distribution}
\end{figure}

\begin{figure*}[htbp]
    \centering
    \includegraphics[width=\textwidth, height=1.2\textwidth]{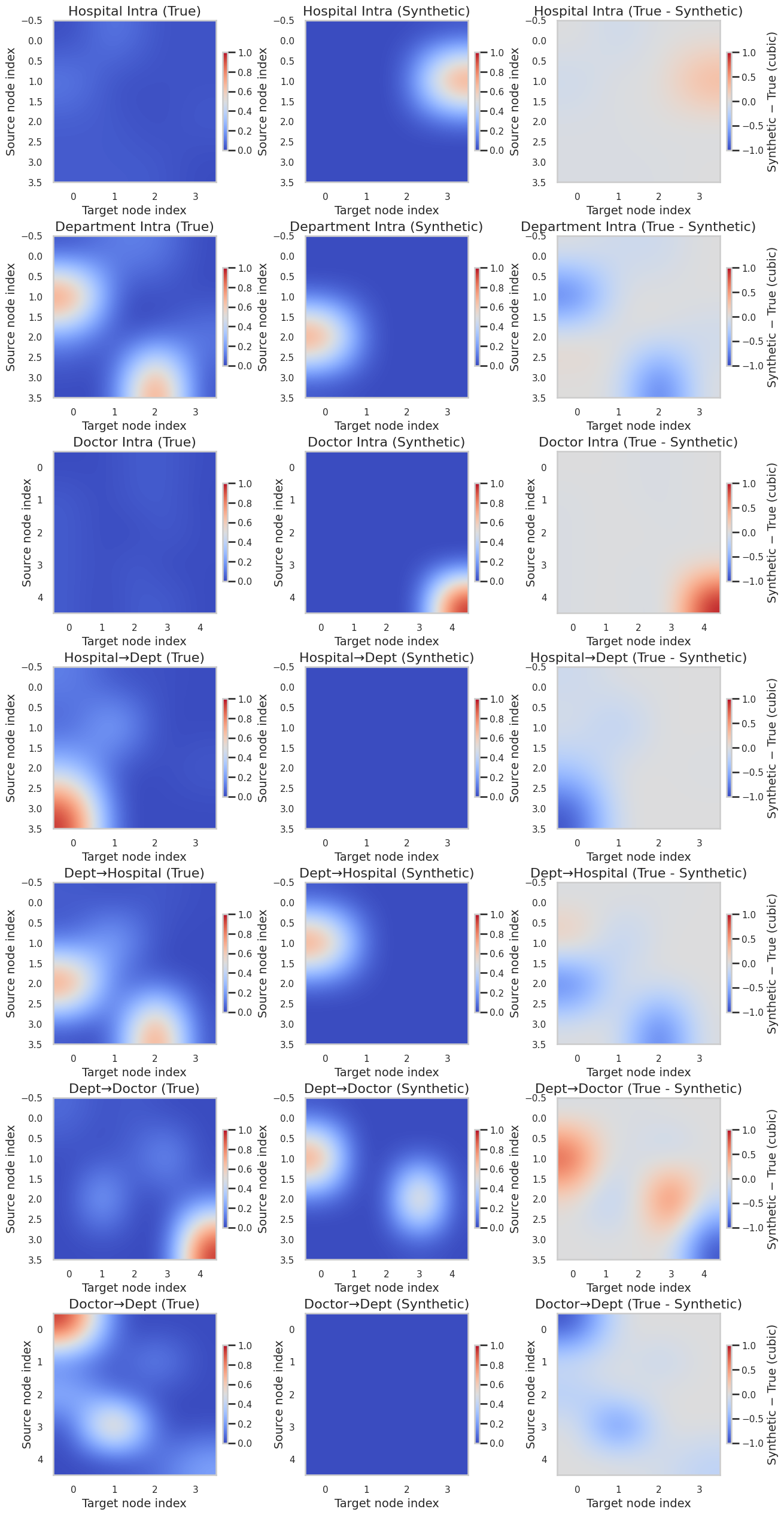}
    \caption{Per-layer \textit{contour} visualizations of the trust matrices with a cubic value transform. 
    Each row corresponds to one network layer: Hospital Intra, Department Intra, Doctor Intra, Hospital$\rightarrow$Department, Department$\rightarrow$Hospital, Department$\rightarrow$Doctor, and Doctor$\rightarrow$Department. 
    Columns show: \textbf{(left)} the true trust matrix $T^{3}$, \textbf{(middle)} the CTGAN–generated synthetic matrix $S^{3}$ , and \textbf{(right)} the signed cubic residual $(S-T)^{3}$ with a zero–centered scale.The horizontal axis is the target (destination) node index, and the vertical axis is the source node index. The first two columns use a 0–1 trust range, while the residual column is symmetrically scaled around zero. }
    \label{fig:contours}
\end{figure*}

 We analysed CTGAN’s fidelity with three complementary diagnostics: (i) elementwise scatter of actual vs.\ synthetic entries (Fig.~\ref{fig:syn_vs_true}), (ii) the global distribution of trust values on a log scale (Fig.~\ref{fig:distribution}), and (iii) per-layer contour plots of $T^{3}$, $S^{3}$, and the signed cubic residual $(S{-}T)^{3}$ (Fig.~\ref{fig:contours}); the cubic transform is purely for visualization. For \emph{Hospitals} and \emph{Doctors}, points concentrate tightly along the $y{=}x$ diagonal, the paired histogram shows aligned non-zero mass and similar tails, and residual energy in the contour rightmost column is minimal. Thus, CTGAN preserves both the marginal distribution of non-zero trust and the relative ordering of social scores, yielding perfect or near-perfect top-$k$ overlap downstream (Figs.~\ref{fig:syn_vs_true}, \ref{fig:distribution}, \ref{fig:contours}).

\par In contrast, the \emph{Department} layer is a tiny (4-node), highly sparse block with many tied rows in $T$. A generator that sensibly concentrates mass near zero (Fig.~\ref{fig:distribution}) can flip marginal ranks, producing diagonal spread in the scatter (Fig.~\ref{fig:syn_vs_true}) and mild, structured residuals in the contours (Fig.~\ref{fig:contours}); rank-based correlations (Spearman/Kendall) may drop even when $\mathrm{P@k}$ remains reasonable. This is a classic small-$n$/heavy-sparsity artifact rather than a modeling failure, and can be mitigated by per-layer generators, reweighting non-zero edges during CTGAN training, or a minor Dirichlet smoothing before row-normalization. Overall, the downstream trust-propagation pipeline remains stable under distributional shift in $\tau$, supporting our generalizability claim (Figs.~\ref{fig:syn_vs_true}, \ref{fig:distribution}, \ref{fig:contours}).

\subsection{Deployment Challenges}  \vspace{.1in}
While our multilayered trust model demonstrates theoretical efficacy, its real-world adoption faces critical challenges: (1) \textbf{Data heterogeneity} between scraped sources (e.g., Practo) and institutional Electronic health records (EHRs) necessitates Fast Healthcare Interoperability Resources (FHIR)-compliant mappings and validation pipelines; (2) \textbf{Temporal rigidity} of static trust scores requires dynamic update mechanisms of trust computation to accommodate dynamic healthcare changes, e.g., doctor transfers, new hospital equipment; (3) \textbf{Scalability limitations} from $O(n^2)$ trust propagation complexity demand distributed graph processing; (4) \textbf{Regulatory constraints}  mandate edge-level differential privacy; (5) \textbf{Clinical interpretability} gaps risk clinician distrust (e.g. reference recommendation of a specific doctor or hospital etc.) without explainable AI frameworks (e.g., SHAP values); and (6) \textbf{Integration barriers} with legacy hospital IT systems may require lightweight API gateways. Addressing these through phased pilot deployments will be essential for operational translation.  
\vspace{.2in}
\section{Conclusion and Future  Work}\label{s:conc} \vspace{.1in}
This paper proposes a novel multi-layered network model for the healthcare system. This model captures the social interactions among fundamental healthcare entities: doctors, departments, and hospitals. The model emphasizes the distinct relationships between these entities, providing a deeper understanding of healthcare system dynamics. Furthermore, this work introduces a trust-based network model that incorporates trust relationships between medical entities.  A `social score' represents the aggregated social importance of each entity within the multi-layered network. This score is calculated by considering both homogeneous (e.g., doctor-doctor interactions) and heterogeneous interactions (e.g., doctor-department interactions).  The social score can be used to distinguish critical medical entities and support the development of recommendation systems and patient support systems. Moreover, in this paper, we validate the proposed model through simulations and demonstrate its usefulness through empirical analysis. It's important to note that while patients are integral to healthcare, modeling patient-patient interactions is considered out of the scope of this work due to potential complexity and minimal impact on the proposed approach. The proposed framework is intended primarily as an operational and decision-support tool for healthcare service optimization, rather than as a clinical or biomedical device validation study. While the model has been empirically validated against real-world ratings of hospitals and departments, future deployments in collaboration with healthcare institutions can incorporate domain-specific workflows and case studies to demonstrate practical impact further.

\par In the future, there are several scopes for extending this work. First, further refinement of trust analysis methods can enhance data reliability. Second, a deeper understanding of potential biases in doctor ratings can improve the model's accuracy. The current trust and social score computation framework, though deterministic, is designed to be compatible with learning-based extensions. Residual social scores and trust matrices can be initialized or updated using machine learning models trained on historical referral patterns, patient outcomes, or other contextual data. Such integration could further enhance adaptability and robustness to heterogeneous and evolving healthcare datasets.
Further, optimizing referral systems within the multi-layered networks using the trust-based approach can lead to more personalized and reliable recommendations for patients. Finally, it may be useful to develop temporal decay mechanisms and robust weighting schemes to mitigate rating biases in trust propagation, enhancing model fairness across dynamic healthcare environments.

\bibliographystyle{elsarticle-num}
\bibliography{Bibliography_HCN}

\end{document}